\documentclass{aa} 

\usepackage{graphicx}
\usepackage{txfonts}
\usepackage[normalem]{ulem}
\usepackage{mhchem}
\usepackage{nccmath}
\usepackage{tabularx}
\usepackage{xspace}
\RequirePackage[english]{babel}
\usepackage{dblfloatfix}

\usepackage[colorlinks=true,linkcolor=blue,citecolor=blue,urlcolor=blue]{hyperref}

\newcommand{\cair} {Ca~II~8542~\AA\xspace}
 
\newcommand{\halpha} {\ce{H\alpha}\xspace}

\def\specand{\,\&\,}        

\def\Hbeta{\mbox{H\hspace{0.1ex}$\beta$}\xspace}

\def\FeI{\mbox{Fe\,I\, 6301\,\AA}\xspace}

\def\CaIR{\mbox{Ca\,II\,\,8542\,\AA}\xspace}

\def\CaIIK{\mbox{Ca\,II\,\,K}\xspace}
\def\CaIIH{\mbox{As}\xspace}
\def\CaIIHK{\mbox{Ca\,II\,\,H{\specand}K}\xspace}

\def\MgIIhk{\mbox{Mg\,II\,\,h{\specand}k}\xspace}

\begin{document}

 \title{Center-to-limb variation of spectral lines and continua observed with SST/CRISP and SST/CHROMIS}

 \author{A.G.M. Pietrow\inst{1,2}
 \and D. Kiselman\inst{1}
 \and O. Andriienko\inst{1}
 \and D.J.M. Petit dit de la Roche\inst{3}
 \and C.J. Díaz Baso\inst{1,4,5}
 \and F. Calvo\inst{1}
     }

 \institute{Institute for Solar Physics, Dept. of Astronomy, Stockholm University, Albanova University Centre, SE-106 91 Stockholm, Sweden
 \and
 Leibniz-Institut für Astrophysik Potsdam (AIP), An der Sternwarte 16, 14482 Potsdam, Germany
 \and
 Observatoire de Gen\`{e}ve, Universit\'{e} de Gen\`{e}ve, Chemin Pegasi 51, 1290 Sauverny, Switzerland
 \and
 Institute of Theoretical Astrophysics, University of Oslo, %
 P.O. Box 1029 Blindern, N-0315 Oslo, Norway
 \and
 Rosseland Centre for Solar Physics, University of Oslo, %
 P.O. Box 1029 Blindern, N-0315 Oslo, Norway
 \\
       \email{apietrow@aip.de}
      }

 \date{Received November 1, 2021; accepted November 25, 2021}

\abstract
{Observations of center-to-limb variations (CLVs) of spectral lines and continua provide a good test for the accuracy of models with a solar and stellar atmospheric structure and spectral line formation. They are also widely used to constrain elemental abundances, and are becoming increasingly more important in atmospheric studies of exoplanets. However, only a few such data sets exist for chromospheric lines.}
{We aim to create a set of standard profiles by means of mosaics made with the CRISP and CHROMIS instruments of the Swedish 1-m Solar Telescope (SST), as well as to explore the robustness of said profiles obtained using this method.}
{For each spectral line, we used a mosaic that ranges from the center to the limb. Each of these mosaics were averaged down to 50 individual spectral profiles and  spaced by 0.02 in the $\mu$ scale. These profiles were corrected for p-mode oscillations, and their line parameters (equivalent width, line shift, full-width at half-maximum, and line depth) were then compared against literature values whenever possible.}
{We present a set of 50 average profiles that are spaced equidistantly along the cosine of the heliocentric angle ($\mu$) by steps of 0.02 for five continuum points between 4001 and 7772 \AA, as well as ten of the most commonly observed spectral lines at the SST (\CaIIHK, H$\beta$, Mg~I~5173~\AA, C~I~5380~\AA, Fe~I~6173~\AA, \FeI, H$\alpha$, O~I~7772~\AA, and \cair).} 
{The CLV of line profiles and continua are shared in the CDS as machine readable tables, providing a quantitative constraint on theoretical models that aim to model stellar atmospheres.}


 \maketitle
%
\section{Introduction}

Observations of solar limb darkening -- or, more generally speaking, the center-to-limb variation (CLV) of continuum radiation and spectral lines -- have been an important tool for probing the solar atmosphere since \citet{Schwarzschild1906} used this phenomenon to show that the photosphere must be in radiative equilibrium \citep[e.g.,][]{hale07, voigt50, Pagel55, horst85, Alissandrakis17}. Perhaps one of the most used limb darkening studies is the one done by \citet{Neckel94}, where the continuum limb darkening was measured for wavelengths between 330 and 1250 nm and presented in the form of fifth-degree polynomials. There are also several other such atlases that focus on various center-to-limb positions, for example the third solar spectrum (SS3)\footnote{Which denotes the ratio between the spectrum at a given $\mu$ position and that at disk center.} by \citet{stenflo15} and \citet{ramelli17}; the CLV line studies by  \citet{David61}, \citet{White62}, \citet{white68}, \citet{Gurtovenko75}, \citet{horst88},  \citet{Ding91}, \citep{Faurobert13}, and \citet{Takeda19}, for example; or the atlas by \citet{Fathivavsari14}.

Center-to-limb variation measurements provide a good test for the accuracy of models of atmospheric structure and spectral line formation. For example, in \citet{unsold55} the  Milne-Eddington model is compared to observations of lines that are dominated by absorption and scattering processes, and similar checks were done more recently by \citet{vernazza76}, \citet{johan18}, and \citet{Ballester21}, for example. Other studies have used such measurements to constrain the elemental abundance of spectral lines \citep[e.g.,][]{moore66, Tiago2009,2021MNRAS.508.2236B}. In these cases, the full-width at half-maximum (FWHM)\ and equivalent width are often reported, or synthesized spectra with different abundance values are compared to the observations. Additionally, studies of the Lyman-$\alpha$ and \MgIIhk lines have been performed by \citet{Bonnet78} and \citet{lemaire81}, for example, and more recently by \citet{gunar20, gunnar21}, and \citet{Rachmeler22} to create quiet-Sun reference profiles which could serve as an incident radiation boundary condition for the radiative transfer modeling of chromospheric and coronal structures. Additionally, CLV measurements of the polarization of these lines have been studied by \citet{Rachmeler22}, full disk integrated atlases were presented by \citet{molaro12}, and the formation height of several chromospheric lines was inferred by means of CLVs by \citet{Wittmann76}.



\begin{table*}[!b]
\centering
\caption{Summary of all data used in this paper, listing the date and time at which they were observed, the campaign to which they belong, the line that was observed, the number of mosaic pointings (nP), the number of wavelength points per scan (n$\lambda$), the cadence (cad), the number of frames per wavelength point (fpw), the central wavelength of each observed line, and the continuum wavelength used for position calibration.}\label{tab:lines}
\begin{tabular}{lllllllll}
\hline\hline

\multicolumn{1}{l}{Observation date, time (UT)} & 
\multicolumn{1}{l}{Campaign} & 
\multicolumn{1}{l}{Line} & 
\multicolumn{1}{l}{nP} & 
\multicolumn{1}{l}{n$\lambda$} & 
\multicolumn{1}{l}{cad [s]} & 
\multicolumn{1}{l}{fpw} & 
\multicolumn{1}{l}{$\lambda_{\rm cent}$ [\AA]}&
$\lambda_{\rm cont}$ [\AA]\\ \hline 

\multicolumn{1}{l}{2019-06-21, 08:23-08:43} &    
\multicolumn{1}{l}{I} &          
\multicolumn{1}{l}{Ca II H 3968 \rm \AA} & 
\multicolumn{1}{l}{24} &          
\multicolumn{1}{l}{27} &          
\multicolumn{1}{l}{20} &          
\multicolumn{1}{l}{25} &          
\multicolumn{1}{l}{3968.47}  &               
4001  \\ \hline               

\multicolumn{1}{l}{2019-06-21, 08:23-08:43} &          
\multicolumn{1}{l}{I} &          
\multicolumn{1}{l}{Ca II K 3934 \rm \AA} & 
\multicolumn{1}{l}{24} &          
\multicolumn{1}{l}{27} &          
\multicolumn{1}{l}{20} &          
\multicolumn{1}{l}{25} &          
\multicolumn{1}{l}{3933.66}  &              
4001 \\ \hline 

\multicolumn{1}{l}{2019-06-21, 08:23-08:43} &          
\multicolumn{1}{l}{I} &          
\multicolumn{1}{l}{Ca II 8542 \rm \AA} & 
\multicolumn{1}{l}{25} &          
\multicolumn{1}{l}{33} &          
\multicolumn{1}{l}{28} &          
\multicolumn{1}{l}{8} &          
\multicolumn{1}{l}{8542.10}  &            
6301 \\ \hline               

\multicolumn{1}{l}{2019-06-21, 08:23-08:43} &          
\multicolumn{1}{l}{I} &          
\multicolumn{1}{l}{Fe I 6301 \rm \AA} & 
\multicolumn{1}{l}{25} &          
\multicolumn{1}{l}{30} &          
\multicolumn{1}{l}{28} &          
\multicolumn{1}{l}{8} &          
\multicolumn{1}{l}{6301.50} &        
6301 \\ \hline        

\multicolumn{1}{l}{2021-06-12, 08:35-09:35} &          
\multicolumn{1}{l}{II} &          
\multicolumn{1}{l}{H{\rm$\alpha$}\: 6563 \rm \AA} & 
\multicolumn{1}{l}{25} &          
\multicolumn{1}{l}{59} &          
\multicolumn{1}{l}{36} &          
\multicolumn{1}{l}{8} &          
\multicolumn{1}{l}{6562.81} &                
4001 \\ \hline 
\multicolumn{1}{l}{2021-06-12, 08:35-09:35} &          
\multicolumn{1}{l}{II} &          
\multicolumn{1}{l}{H{\rm$\beta$}\: 4861 \rm \AA} & 
\multicolumn{1}{l}{25} &          
\multicolumn{1}{l}{30} &          
\multicolumn{1}{l}{36} &          
\multicolumn{1}{l}{15} &          
\multicolumn{1}{l}{4861.35 } &           
4001 \\ \hline   

\multicolumn{1}{l}{2021-06-12, 08:35-09:35} &          
\multicolumn{1}{l}{II} &          
\multicolumn{1}{l}{Mg I 5173 \rm \AA} & 
\multicolumn{1}{l}{25} &          
\multicolumn{1}{l}{73} &          
\multicolumn{1}{l}{23} &          
\multicolumn{1}{l}{8} &          
\multicolumn{1}{l}{5172.68}&                
4001 \\ \hline 

\multicolumn{1}{l}{2021-06-12, 09:41-10:26} &          
\multicolumn{1}{l}{II} &          
\multicolumn{1}{l}{C I 5380 \rm \AA} & 
\multicolumn{1}{l}{25} &          
\multicolumn{1}{l}{57} &          
\multicolumn{1}{l}{13} &          
\multicolumn{1}{l}{8} &          
\multicolumn{1}{l}{5380.34}  &               
5382\\ \hline

\multicolumn{1}{l}{2021-06-19, 10:03-10:31} &          
\multicolumn{1}{l}{II} &          
\multicolumn{1}{l}{Fe I 6173 \rm \AA} & 
\multicolumn{1}{l}{25} &          
\multicolumn{1}{l}{27} &          
\multicolumn{1}{l}{7} &          
\multicolumn{1}{l}{8} &          
\multicolumn{1}{l}{6173.33} &               
6172 \\ \hline  

\multicolumn{1}{l}{2021-06-19, 10:40-11:01} &          
\multicolumn{1}{l}{II} &          
\multicolumn{1}{l}{O I 7772 \rm \AA} & 
\multicolumn{1}{l}{25} &          
\multicolumn{1}{l}{23} &          
\multicolumn{1}{l}{7} &          
\multicolumn{1}{l}{8} &          
\multicolumn{1}{l}{7771.94}&                
7771 \\ \hline 

\end{tabular}

\end{table*}


Typically, CLV observations are recorded in one of three ways.
\begin{enumerate}
\item One can place a slit or imager beyond the western limb of the Sun and turn off the telescope tracking. This causes the Sun to drift past the field of view (FOV) at a constant speed, and the resulting signal is then recorded \citep[e.g.,][]{Wittmann76, Neckel94, mats16}. This method requires many full disk scans in order to average out the small-scale structure across the solar diameter, as well as the absence of any larger structures that could contaminate the average. These kinds of measurements are typically done from west to east, but they can be achieved in other directions by modifying the tracking speed. This method is restricted to monochromatic measurements or spectrographs that give simultaneous data at several wavelengths.

\item One can observe at several distinct disk positions \citep[e.g.,][]{Ding91,Pierce82, Tiago2009, Lind17}. This method is easier to work with because the observer can choose the location without activity on the Sun for each $\mu$ (the cosine of the heliocentric angle). Additionally, the fact that tracking is active allows for each observation to stay pointed at one location for a longer time to form temporal averages. This comes at the cost of precision, as there is typically no overlap between consecutive pointings. 

\item One can create mosaics of monochromatic images (or sets of monochromatic images) where large amounts of pixels can be averaged, potentially excluding areas that contain active regions \citep{gunnar21}. This method allows the user to choose between longer integration times per position at the expense of time or vice versa. This is the approach taken here.
\end{enumerate}

In this study we present a quiet-Sun CLV analysis of the \CaIIHK, \halpha, \Hbeta, Mg I 5173 \AA, and \cair lines. These are strong, broad spectral lines with their cores having been formed in between the temperature minimum and the upper chromosphere. In addition to that, we also study the C I 5380 \AA, Fe I 6173 \AA, \FeI, and O~I~7772~\AA\  lines which are mainly formed in the photosphere. We also observe the Fe I 5379 \AA, Fe~I~5172~\AA, Ti~II~5381~\AA, and Ti~I~5174~\AA~ lines around the C I 5380 \AA~ and Mg I 5173 \AA~ lines, but we do not study them.

We carried out these observations with a series of semi-overlapping mosaics acquired along a solar radius. The main goal of this study is to present representative profiles of the lines mentioned above, as well as to test their robustness by comparing the spectral parameters extracted from these lines against the ones available in the literature. We expect that these profiles will be useful for abundance studies as well as reference profiles, and as approximate background profiles in the study of optically thin chromospheric structures as was done using a cloud model in \citet{pietrow22}. Additionally, we aim to use our limb darkening curves to improve the intensity calibration of the \citep[SSTRED;][]{jaime15, mats21} data-processing pipeline of the Swedish 1m-Solar Telescope \citep[SST][]{Goran2003sst,Goran2003ao} .

\section{Observations and data processing}\label{observations}
\begin{figure}
  \centering
  \includegraphics[width=0.48\textwidth]{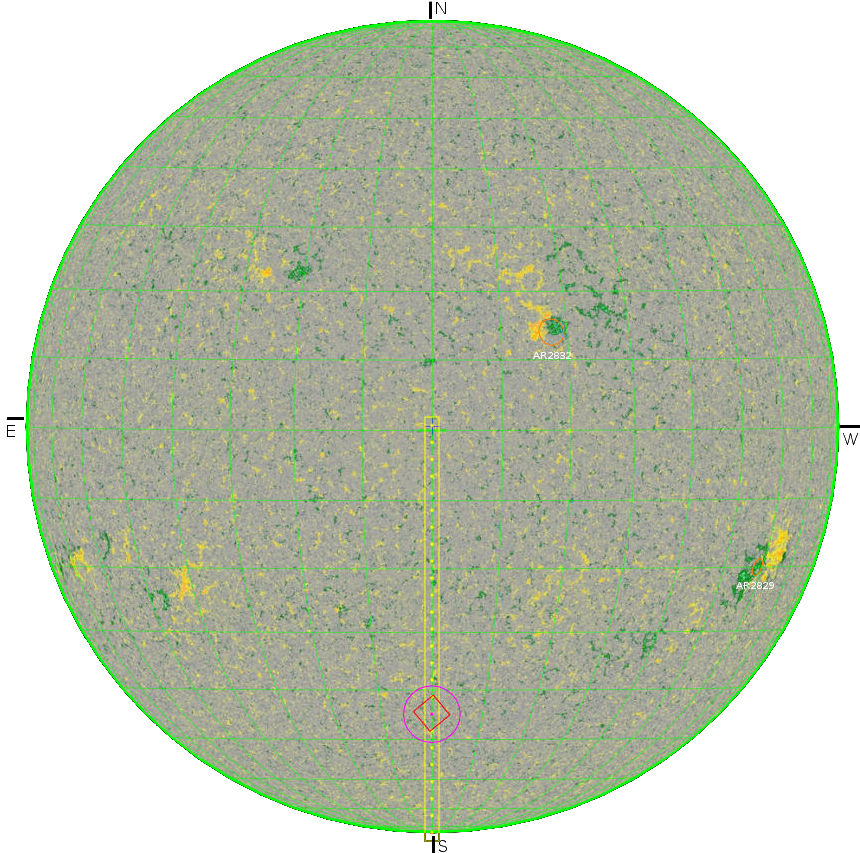}
  \caption{Screen capture of the primary image guider of the SST showing the HMI magnetogram with 25 pointings in a mosaic that spans from the solar south pole to the disk center, taken on 12 June 2021. The red marker represents the location at which the telescope was aimed at the moment that the image was taken.}
  \label{fig:pig}
\end{figure}

\begin{figure*}
\centering
\includegraphics[width=\textwidth]{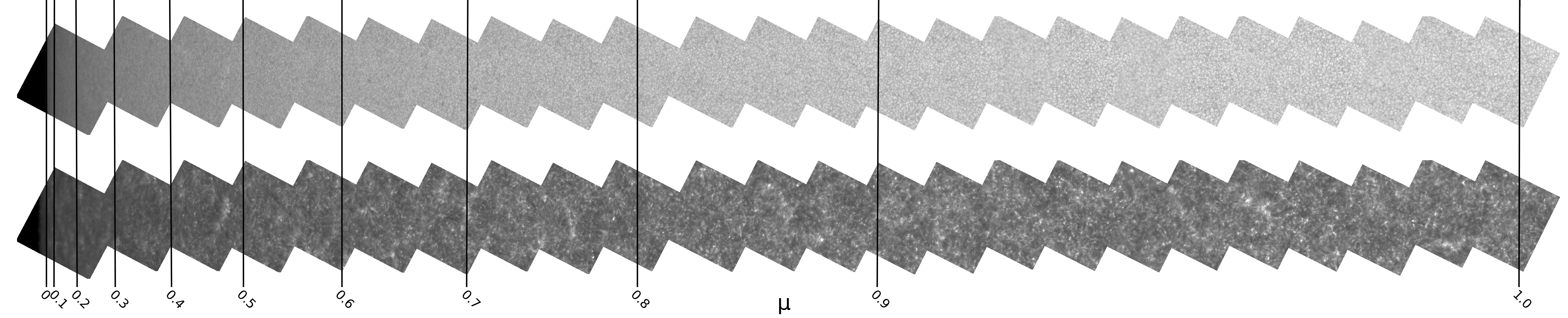}
\caption{Overview of mosaics taken during Campaign I, starting from the solar limb on the left and going to disk center on the right. The mosaics span roughly 1000\,$\times$\,80 arcseconds \textbf{Top:} Continuum near the blue wing of \FeI. \textbf{Bottom:} Line center of \CaIR. It can be seen that the Sun seems to extend past $\mu=0$ in the chromosphere.}
\label{fig:overview}
\end{figure*}

Our data consist of ten mosaics, each spanning one solar radius, taken with the SST, using both the CRisp Imaging SpectroPolarimeter \citep[CRISP,][]{Scharmer08} and the CHROMospheric Imaging Spectrometer \citep[CHROMIS;][]{Scharmer17} simultaneously. All mosaics were observed during two campaigns, one in June 2019 and another in June 2021, which henceforth are called Campaign I and Campaign II, respectively. Observations from Campaign I were taken from the disk center toward the solar north pole, while observations from Campaign II were taken from the solar south pole toward the disk center (see Fig.~\ref{fig:pig}). In all cases, there was roughly a 30\% area overlap between each consecutive pointing (see Fig.~\ref{fig:overview}).

To maximize the number of quiet-Sun pixels in each FOV, all observations were carried out during periods when there were no active regions along the mosaic path. The observations were carried out under variable seeing conditions. This is acceptable because we are primarily interested in taking large-scale averages along the mosaic and using the fine-scale structure solely for alignment purposes. The spectral lines from Campaign I were observed with one spectral scan per pointing during better seeing, while the lines from Campaign II were taken with three spectral scans per pointing in order to increase the likelihood of a good scan during poor seeing conditions. All the spectral lines were observed without polarimetry, except for \FeI.

We list each of the observed lines in Tables~\ref{tab:lines} and \ref{tab:linesteps}. All wavelength positions are given relative to their line center. We note that not all lines are symmetrically sampled; in such a case, a value only has a $+$ or $-$ instead of a $\pm$ symbol in front of it. 

\begin{table}[]
\centering
\caption{FPI tuning positions for each observed line.}
\begin{tabularx}{\columnwidth}{X}
\hline
\hline
Ca II K 3934 \AA \\ \hline
$\pm$1495, $\pm$1170, $\pm$910, $\pm$650, $\pm$585, $\pm$520, $\pm$455, $\pm$390, $\pm$325, $\pm$260, $\pm$195, $\pm$130, $\pm$65, and 0 m\AA.\\ \hline
\hline
Ca II H 3968 \AA \\ \hline
$-$1235, +1040 $-$845, $\pm$650, $\pm$585, $\pm$520, $\pm$455, $\pm$390, $\pm$325, $\pm$260, $\pm$195, $\pm$130, $\pm$65, and 0 m\AA.  \\ \hline
\hline
H$\rm\beta\:$ 4861 \AA \\ \hline
$\pm$600, $\pm$540, $\pm$480, $\pm$420, $\pm$360, $\pm$300, $\pm$240, $\pm$180, $\pm$120, $\pm$60, and 0 m\AA.  \\ \hline
\hline
Mg I 5173 \AA \\ \hline
$-$1540, $\pm$1485, $\pm$1430, $\pm$1375, $\pm$1320, $\pm$1265, $\pm$1210, $\pm$1155, $\pm$1100, $\pm$1045, $\pm$990, $\pm$935, $\pm$880, $\pm$825, $\pm$770, $\pm$715, $\pm$660, $\pm$605, $\pm$550, $\pm$495, $\pm$440, $\pm$385, $\pm$330, $\pm$275, $\pm$220, $\pm$165,  $\pm$88, $\pm$77,  $\pm$66,  $\pm$55,  $\pm$44,  $\pm$33,  $\pm$22,  $\pm$11, and  0 m\AA. \\ \hline
\hline
C I 5380 \AA \\ \hline
$-$1235, $-$1178, $-$1121, $-$1064, $-$1007, $-$950, $-$893, $-$836, $-$779, $-$722, $-$665, $-$608, $-$551, $-$494, $-$437, $-$380, $-$323, $-$266, $-$209, $-$152, $-$133, $-$114,  $-$95,  $-$76,  $-$57,  $-$38,  $-$19, 0,  19,  38,  57,  76,  95,  114,  133,  152, 171,  190,  247,  304,  361,  418,  475,  532,  589, 646,  703,  760,  817,  874,  931,  988, 1045, 1102, 1159, 1216, and 1273 m\AA. \\ \hline
\hline
Fe I 6173 \AA  \\ \hline
$-$960, $\pm$840, $\pm$720, $\pm$600, $\pm$480, $\pm$360, $\pm$144, $\pm$120, $\pm$96, $\pm$72, $\pm$48, $\pm$24, and  0 m\AA. \\ \hline
 \hline
Fe I 6301 \AA \\ \hline
$-$1270, $-$1235, $-$1200,$-$1165, $-$1130, $-$1095, $-$1060, $-$1025, $-$990, $-$955, $-$920, $-$885, $-$850, $-$815,$-$780, $-$745, $-$710, $-$240, $-$210, $-$180, $-$150, $\pm$120, $\pm$90, $\pm$60, $\pm$30, and 0 m\AA. \\ \hline
 \hline
H$\rm\alpha\:$ 6563 \AA \\ \hline
$\pm$2015, $\pm$1860, $\pm$1705, $\pm$1550, $\pm$1395, $\pm$1240, $\pm$1085, $\pm$930, $\pm$775, $\pm$620, $\pm$589, $\pm$558, $\pm$527, $\pm$496, $\pm$465, $\pm$434, $\pm$403, $\pm$372, $\pm$341, $\pm$310, $\pm$279, $\pm$248, $\pm$217, $\pm$186, $\pm$155, $\pm$124,  $\pm$93, $\pm$62,  $\pm$31, and   0 m\AA. \\ \hline
 \hline
O I 7772 \AA \\ \hline
$\pm$980, $\pm$735, $\pm$392, $\pm$343, $\pm$294, $\pm$245, $\pm$196, $\pm$147,   $\pm$98,   $\pm$49,     and 0 m\AA. \\ \hline
\hline
Ca II 8542 \AA \\ \hline
$\pm$1815, $\pm$825, $\pm$770, $\pm$715, $\pm$660, $\pm$605, $\pm$550, $\pm$495, $\pm$440, $\pm$385, $\pm$330, $\pm$275, $\pm$220, $\pm$165, $\pm$110, $\pm$55, and 0 m\AA. \\ \hline

\end{tabularx}\label{tab:linesteps}

\end{table}


The SSTRED pipeline \citep{jaime15,mats21} has been designed to process the data from the SST. It not only includes dark, flat field, and polarimetric calibrations, but it also performs image restoration, removing optical aberrations caused by turbulence in the atmosphere (and partially corrected for by the SST adaptive optics) using Multi-Object Multi-Frame Blind Deconvolution \citep[MOMFBD;][]{mats02,vanNoort05}. The MOMFBD algorithm assumes that the point spread function (PSF) is constant within subfields of ~5 arcseconds squared. The restored subfields are then mosaicked, resulting in a restored version of the entire FOV. MOMFBD methods work best when the contrast is high, the noise is low, and the exposure time is short. However, under poor seeing conditions (Fried parameter $r_0$ of 5~cm or lower), the reconstruction can fail. This is characterized by the subfields becoming apparent as a checkered pattern of squares in the images. In this situation, the intensity may not be conserved, and thus the resulting spectra are expected to become affected and are unusable for the current purpose.

To overcome this problem under poor seeing conditions, we used a modified version of the SSTRED pipeline that performs all the data processing but does not apply MOMFBD. We believe that this 
can also be useful for time series with seeing dips as they allow for the temporal continuity of the data to be maintained, for data where spatial resolution is not relevant, and as a quick look tool. This feature has now been implemented into the SSTRED pipeline by a keyword that allows the user to set the amount of MOMFBD iterations to 0, thus only skipping the reconstruction while keeping the rest of the process. 

The SSTRED pipeline makes use of the flat fields to produce a so-called cavity map that describes the variations of the central wavelength of the CRISP and CHROMIS instrumental profile over the FOV. In order to avoid unnecessary interpolation in the data, SSTRED does not apply the cavity-map corrections by default, but it leaves it to the end-user to apply it –– which is typically done by correcting derived Doppler shifts. Since the large-scale variations gave a strong imprint on the derived CLV curves, even after the spatial averaging, we applied the cavity-map correction by interpolating the intensities for each pixel between different wavelength frames. In the case of spectroscopic inversions, the cavity map is often subtracted from the resulting velocities in order to correct for the line shifts without changing the profiles beforehand \citep[e.g.,][]{Carlos19,pietrow2020}.


\section{Methods}\label{methods}
In this section we describe the methods used for the alignment of the mosaics, intensity, and position calibration of the data and the removal of p modes.
\subsection{Mosaic alignment}
The SST pointing and tracking system has features to simplify the making of mosaics. These features have allowed studies such as \citet{Scullion2014} where a large active region was observed in a 9\,$\times$\,6 pointing mosaic that spanned a FOV of 280\,$\times$\,180 arcseconds. \citet{Adur2020} performed a partially overlapping mosaic with 17 pointings of the north pole region, and \citet{Gregal21} presented a smaller 2\,$\times$\,2 mosaic of a flaring region that spanned a FOV of roughly 110\,$\times$\,110 arcseconds. Our mosaics, which look at the quiet Sun, consist of 1\,$\times$\,25 pointings and span a FOV of roughly 60\,$\times$\,1000 arcseconds. 

Typically, mosaics of this type are aligned by using a cross-correlation algorithm \citep[e.g.,][]{reardon2012,hammerschlag2013}. However, these algorithms require several distinguishing features to operate, which for quiet-Sun observations is not the case, especially very close to the limb where the image contrast decreases. Therefore, the images were aligned by manually looking for matching patterns in the granulation and bright points in the overlapping areas between pointings, as well as with SDO/HMI continuum images \citep{SDO2012} which were spatially enhanced  \citep{Carlos2018}. Afterward, the alignment was refined by means of an ISPy mosaic alignment tool created for this project \citep{ISPy2021}. This tool allows the user to plot two consecutive mosaic pointings at an approximate location,  which are moved, blinked, and further aligned.

A perfect match is not possible since our images are not destreched and taken under suboptimal seeing conditions. For this reason, we assume a conservative estimate for the alignment error by assuming that the pointing is accurate to within 16 pixels, which corresponds to roughly 1 and 0.5 arcseconds for CRISP and CHROMIS, respectively. Finally, we obtained our absolute solar coordinates with respect to the disk center from the alignment of our data with the enhanced SDO/HMI continuum observations. 

\subsection{Intensity calibration}
The integration time is constant, which means that apart from the varying extinction, there is a linear relation between detector counts and solar intensity. We corrected our data for the atmospheric extinction variations, which are caused by the change in the Sun's altitude during the acquisition of the mosaic, by regularly taking a disk center observation in between our observations. In Campaign I we took one disk center observation before starting the mosaic and one at the end. In Campaign II we added a third disk center observation halfway through the mosaic. A quadratic function was then fitted to these observations to obtain an approximation for the extinction at the time of the other pointings. This method has also been integrated into the SSTRED pipeline, while before it was only correcting for a single extinction value. After correcting for the extinction, we calibrated the intensity of all pointings with respect to the disk center pointing by simultaneously fitting several wing points of the averaged disk center profile to the solar atlas by \citet{Neckel1984}. We found that this works well and that no additional calibration between frames are necessary in order to assure a smooth transition between each overlapping frame. We do note that this method cannot be assumed to work with rapidly varying extinction, caused by clouds or calima for example.

\subsection{Position calibration}
For each mosaic, we used the Pythagorean theorem to calculate the distance of each pixel in the mosaic from the disk center, which can be used to collapse the mosaic into a 1D intensity array. We defined the limb in the same way as \citet{Neckel94}, which is the inflection point of the sharp tail end of the intensity array in the continuum. This was done by taking the numerical derivative of a single-pixel column from the disk center to the limb, which allowed us to accurately pinpoint the inflection point, which we defined as $\mu=0$, to within 3 pixels (about 0.2\arcsec) for CRISP). Pixels beyond this point were discarded.

The continuum wavelengths used for each spectral line are given in Table~\ref{tab:lines}. The broad lines do not have access to a continuum within their prefilter range, and therefore a compromise was needed. The largest wavelength difference between a line and the continuum wavelength used to define the limb position is for \halpha, which uses the CHROMIS continuum filter at 4001\,\AA.

The wavelength difference between the continuum-defined wavelength and the chromospheric wavelength that it is applied to in itself is not expected to give a large error. \citet{Neckel94} reported that the position of the limb varies by 0.12 arcseconds between 3010 and 10990 \AA, and up to 0.5 arcseconds for bad seeing at 400 nm (being smaller for longer wavelengths). In addition to that, the atmospheric differential refraction seemingly compresses the Sun along one axis. However, we estimate that the difference in apparent elevation of the solar limb in our observations that are caused by this effect should be less than 1\arcsec\ in the extreme case of \halpha and 4001\,\AA. With this in mind, and considering seeing-induced smearing, we chose a limb distance of 3\arcsec\ (corresponding to $\mu=0.08$) as a limit beyond which our results for the broad lines should be taken with caution.

The commonly cited CRISP and CHROMIS image scales of 0.058"/pixel and 0.0375"/pixel, respectively, were derived from measurements by \citet{noren}. These values are accurate enough when working inside a single FOV or a small mosaic, but not over the range of 25 pointings. Therefore, we redefine the angular scale for our 1D center-to-limb curves to be equal to the angular radius of the Sun as viewed from Earth for the date of observation. For campaigns I and II, this is 944 and 945 arcseconds, respectively. Here we find image scales of 0.0584"/pixel and 0.0384"/pixel for the two instruments, respectively. Based on these scales, we can assign a $\mu$ value at each pixel in the mosaic. Finally, we binned down the data into 50 bins that each span $\Delta_\mu = 0.02$ in the $\mu$ scale.

\subsection{Removing p-mode oscillations}
\begin{figure*}
          \centering
          \includegraphics[width=\textwidth, trim={4.8cm 0.5cm 4cm 1cm},clip]{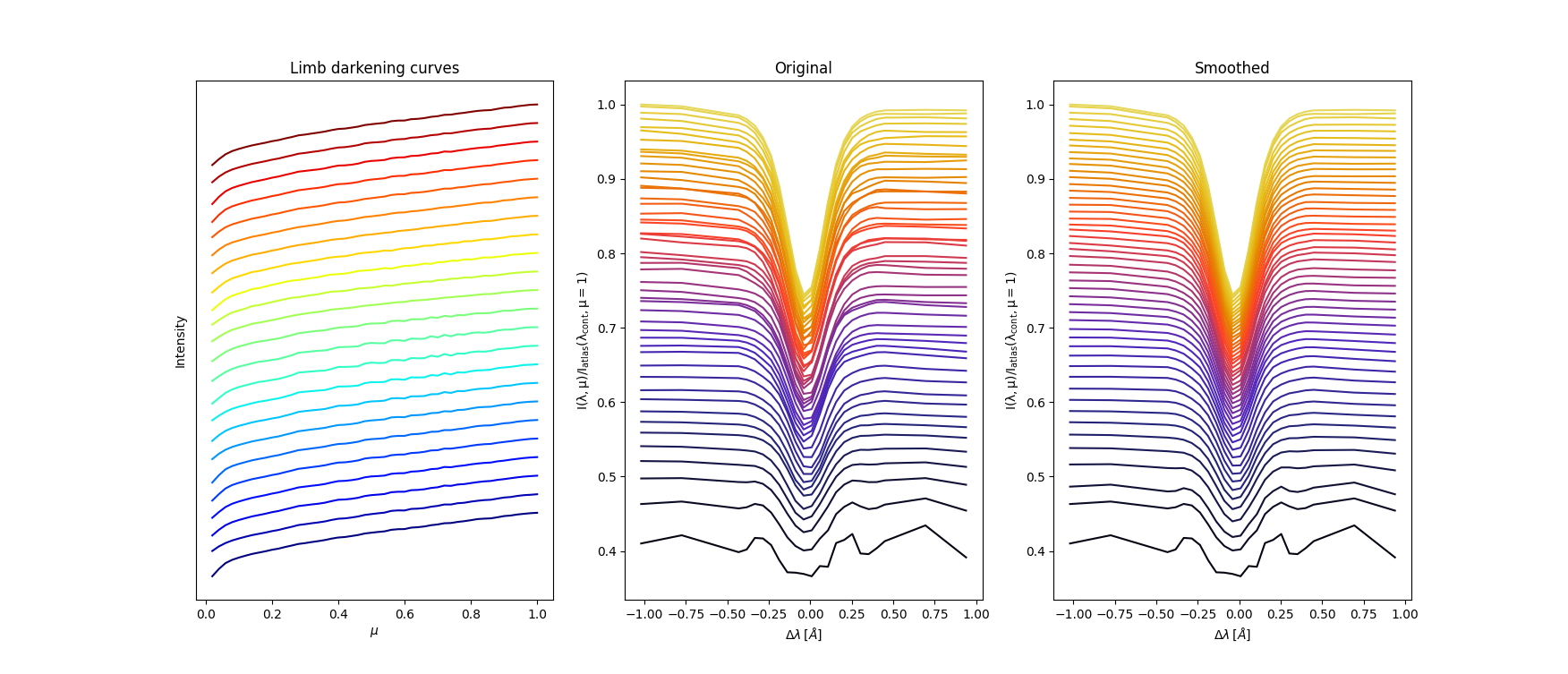}
          \caption{Comparison of normalized average O~I~7772~\AA\ profiles before and after smoothing to remove effects from p modes. \textbf{Left:} Un-smoothed limb darkening data of each wavelength point in the O~I~7772~\AA\ line on an arbitrary intensity scale. The curves are ordered by wavelength, with the top (red) curve representing the red continuum, the middle (green) curve representing the line core, and the bottom (blue) curve representing the blue continuum. The oscillatory behavior is most pronounced around the line core (green), between a $\mu$ of 0.3 to 0.8. \textbf{Middle:} Original representative profiles, ranging from $\mu=1$ to $\mu=0.02$. The profile representing $\mu = 1$ is shown in yellow at the top of the figure, with each consecutive profile beneath it in a different color. \textbf{Right:} Same as the middle panel, but with most of the p-mode effects removed after limb darkening curves were smoothed. }
          \label{fig:pmodes}
\end{figure*}

After the examination of the limb darkening curve of each wavelength point, we found that our observations are affected by 5-minute oscillations (p modes, \citealt{Leighton60}). The oscillations in intensity as a function of $\mu$\, caused by this effect become especially pronounced in the steep parts of the line wings, where a small shift in wavelength becomes a relatively large change in intensity. Typically, p-mode oscillations are removed by averaging over time. Since we do not have a long enough time series for these mosaics, we instead chose to work with the limb darkening curves in the $\mu$ scale. This was done by applying a standard box filter, also known as a moving average,  with a width of $5\Delta_\mu$ points to the central part (excluding the three outermost points on both sides) of each limb darkening curve (in $\mu$ scale) in order to preserve the edges of the limb darkening curves. The smoothed curves are illustrated in the left panel of Fig.~\ref{fig:pmodes}, while the profiles before and after the smoothing are shown in the two remaining panels. We use both these smoothed $\mu$-averaged profiles as well as the original profiles for all of the following steps in this work and include them in the supplemental material.

\section{Results and discussion}\label{results}

We created a set of 50 averaged line profiles for each spectral line along the observed mosaic. Each profile represents a $\mu$ position between 0 and 1, spaced by the step of $\Delta_\mu = 0.02$ This means that while each of our profiles represents an equal distance in the $\mu$ scale, they span uneven distances in the angular scale, with the first bin spanning nearly 200 arcseconds and the last one less than one arcsecond (See Fig. \ref{fig:overview}). However, due to the spatial resolution of CRISP and CHROMIS, this means that even the smallest bins still contain an average of hundreds of pixels (see Fig.~\ref{fig:overview}).

\begin{figure*}
          \centering
          \includegraphics[width=0.80\textwidth, trim={1cm 2cm 0.5cm 2cm},clip]{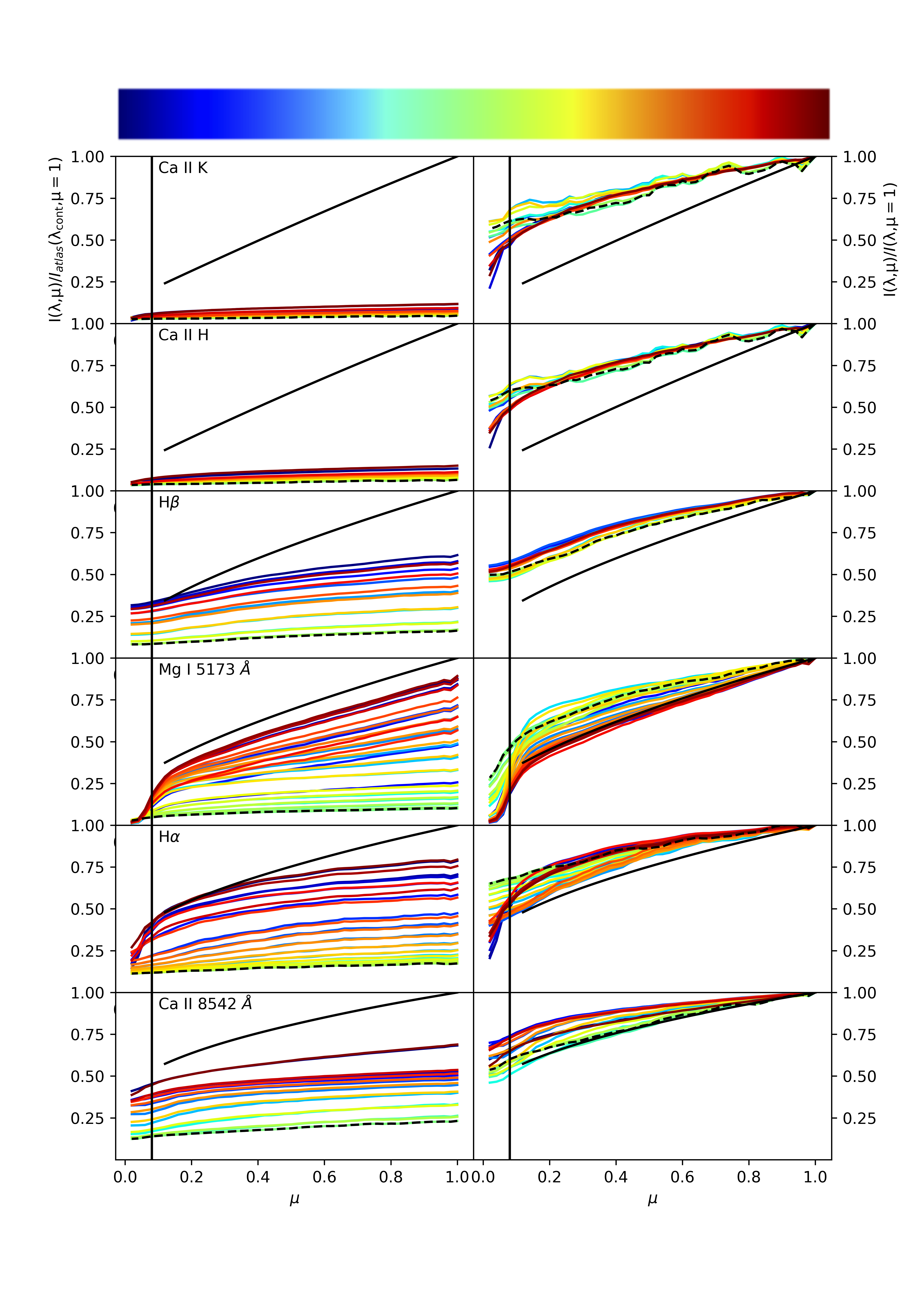}
          \caption{CLV curves of our chromospheric lines ordered by wavelength. In the left column, each CLV curve is shown in atlas calibrated intensities, in the right column each curve is normalized to one to better illustrate the differences in gradient of each wavelength point. The colors of the lines show their approximate wavelength position within the profile, with blue being the blue wing, green the line core, and red the red wing (see colorbar above the plots). The continuum intensity by \citet{Neckel94} is denoted by a solid black line, and the line core by a dashed black line. Figure~\ref{fig:chrom2} shows an enlarged version of the first two panels in the left column. We only plot every other wavelength point to preserve clarity in the figure. The black line at $\mu=0.08$ indicates the point to which we trust the accuracy of our data. }
          \label{fig:chrom}
\end{figure*}

\begin{figure*}
          \centering
          \includegraphics[width=0.95\textwidth]{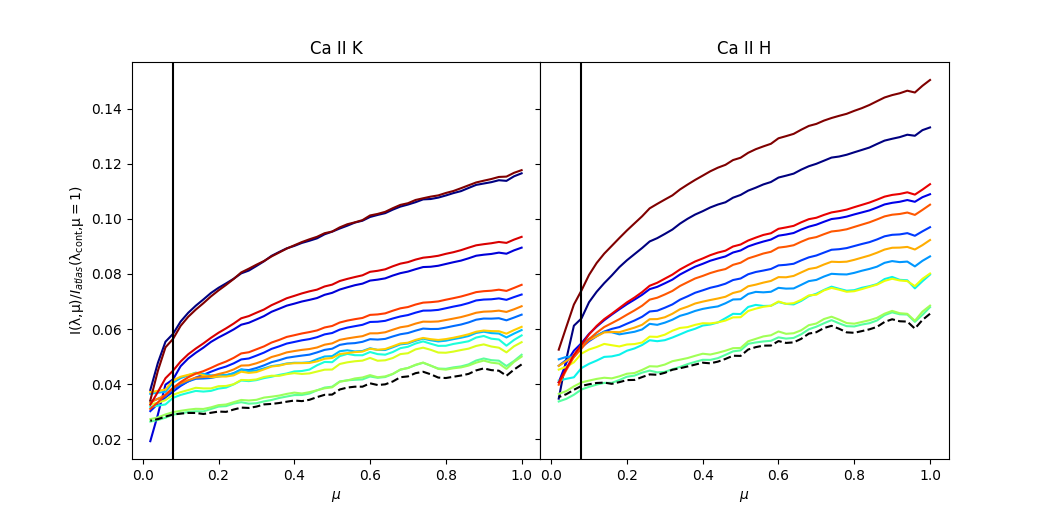}
          \caption{Enlarged version of the first two panels in the left column of Fig.~\ref{fig:chrom}.}
          \label{fig:chrom2}
\end{figure*}

Figures \ref{fig:chrom} and \ref{fig:chrom2} show the CLV curves of the selected wavelength points of the six broad lines. The curves are plotted in two ways: with calibrated atlas intensities in the left column and normalized to the intensity at $\mu=1$ in the right one. The color of the lines varies as shown in the colorbar above the images, with wavelength points on the blue side of the line core being drawn in blue, and red ones in red. Additionally, we have marked the line center with a black dashed curve and the line continuum as given by \citet{Neckel94} with a solid black line. We discuss these plots further in section \ref{broadsec}.

In order to compare our data to other observations, we calculated a set of parameters for each $\mu$ position of each spectral line. For photospheric lines, these are the equivalent width (W$_\mu$), the shift of the line center relative to its position at the disk center, the FWHM, and the line depth. We define the equivalent width for a discretely sampled spectral line as follows, 
\begin{ceqn}
\begin{align}
          W_\mu=\sum^{n\lambda}_{i=0} (1-  I(\lambda_i,\mu)/I_{\rm c}) \Delta\lambda_i \ ,
\end{align}
\end{ceqn} 
where $I(\lambda_i,\mu)/I_{\rm c}$ is the intensity at a given wavelength point normalized to the continuum intensity $I_{\rm c}$, $n\lambda$ is the number of wavelength points, and $\Delta\lambda_i$ is the spacing between each consecutive point. 

Our set of broad spectral lines cannot be fully captured within the prefilter range of the instruments. For these lines, we replaced the equivalent width and the FWHM with the "bound equivalent width" (WB$_\mu$) and the core width (CW), respectively. For the latter, we followed a convention introduced by \citet{2009cauzzi}. This modified FWHM-like parameter is taken halfway between the average intensity of both sides of the profile at a given separation parameter $\Delta\lambda_l$ and the line core. In \citet{2009cauzzi}, the values for \halpha and \cair were chosen to be 900 and 600 m\AA,\ respectively, based on a large-scale inspection of spectral profiles where the boundary between the chromospheric core and the photospheric wings was determined. We adopt these values and define our own for the remaining broad lines in Table~\ref{tab:width}. This allows us to study the behavior of the line cores without being affected by the broad photospheric wings.  

For \CaIIHK we needed yet another approach because of the highly irregular shape of the profiles. Instead of the equivalent width and FWHM, we show the relative shift of the K2 and H2 peaks compared to the line center and their separation as a function of $\mu$, respectively. These peaks can be difficult to find in quiet-Sun profiles, especially the red peak, which tends to disappear in the averaged profiles \citep[e.g.,][]{Druzhinin87}. However, we can still see a clear imprint of both peaks in the wavelength-dependent standard deviation of each averaged profile, which allows us to estimate their location within 50 m\AA. We refer the reader to \citet{Linsky70}, \citet{Linsky17}, and \citet{Ayers19} for a comprehensive review about the Ca II H \& K lines, although the latter two citations  primarily focus on the K line so as to avoid effects caused by the H$\epsilon$ line.

All of these parameters are evaluated by fitting a univariate spline to each of the profiles. This is done to preserve any asymmetries in the lines that would be lost by fitting a Voigt or similar function. This method breaks down around $\mu$ = 0.2, when the continuum intensity starts to decrease rapidly and most lines start showing an emission-line-like behavior. This loss of smoothness in the line profile increases the degeneracy of possible spline fits, most of which have sharp peaks that greatly affect the inferred parameters. In addition, when the viewing angle is almost parallel to the solar surface, the averages contain a smaller number of pixels, so the variations observed in the spectral shapes are subsequently dominated by specific features (such as spicules) and to achieve a reliable average we would need many observations of these regions \citep{hiva17}. Therefore, we limit our study to $\mu$ = 0.2 and above in order to avoid introducing spurious results.

\subsection{Continuum measurements}

\begin{figure}
          \centering
          \includegraphics[width=0.9\columnwidth, trim={1.1cm 1.0cm 1.7cm 1.2cm},clip]{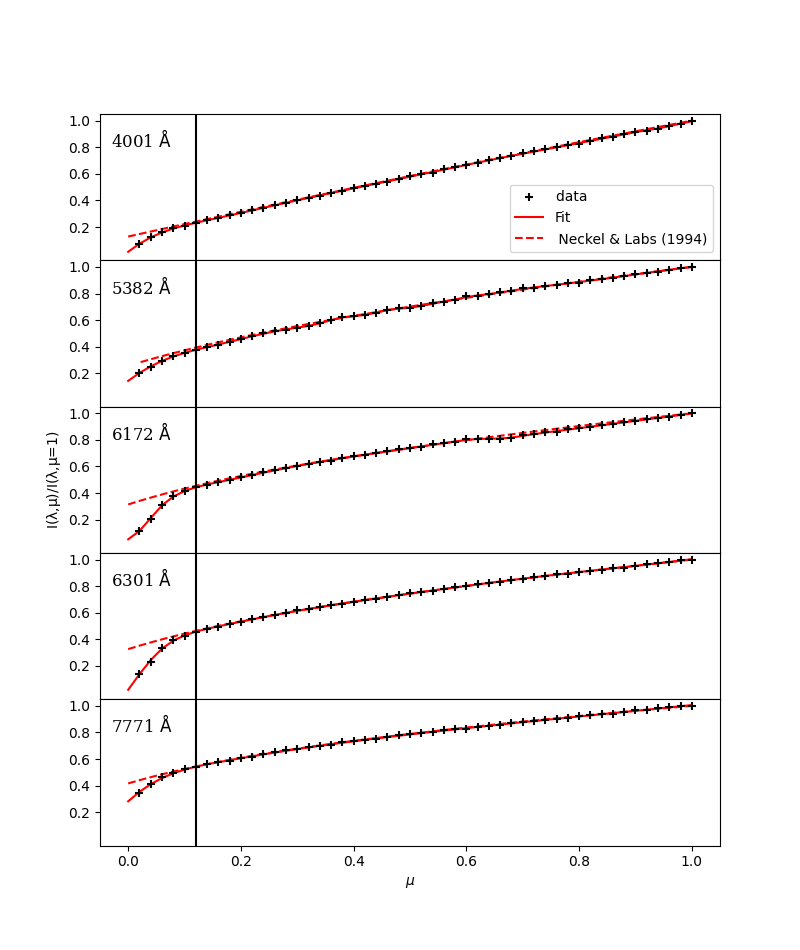}
         \caption{Limb darkening as a function of $\mu$ of five continuum points for our mosaic, plotted on top of the predicted limb darkening curve from \citet{Neckel94} (dashed red) and a univariate spline fitted to the data (red).  The vertical black line shows the furthest measurement toward the limb done by \citet{Neckel94}.}
          \label{fig:continuum}
\end{figure}

In order to validate this method versus more classical approaches, we start with a comparison between the continuum limb darkening given by \citet{Neckel94} and our observations. For this purpose, we use the continuum points at 4001, 5382, 6172, 6301, and 7771 \AA. In Fig.~\ref{fig:continuum} we see that our data agree with the literature values for values of $\mu$ larger than 0.12, while below it our values go down steeper. This is due to the fact that \citet{Neckel94} excluded data taken within 7 arcseconds of the limb to avoid seeing related uncertainties. This corresponds to a $\mu$ of roughly 0.12 when using their solar radius value. This means that our data are in full agreement with \citet{Neckel94}, but that their curves should not be used beyond the aforementioned limit. \citet{neckel03,neckel05}  suggests that the extrapolated values from the \citet{Neckel94} data at $\mu$ = 0.0 give a meaningful "limb temperature" of 4546~K that is nearly constant over wavelength. This is not the case with our extrapolated curves, which give values between roughly 3000 and 4300~K for the five continuum points that we measured. However, it is important to note that the extrapolation of our fitted spline is very sensitive to the three points closest to the limb, and that changing the intensity of any of these three points by as little as 5\% changes the resulting brightness temperature by 200~K.

The uncertainties on these curves are dominated by the 5-minute oscillations, as the classical uncertainties become very small due to the large number of pixels over which each point is averaged. Therefore, we estimate an uncertainty of these values of 1\%.

\begin{figure}
          \centering
          \includegraphics[width=0.9\columnwidth, trim={1.1cm 1.0cm 1.7cm 1.2cm},clip]{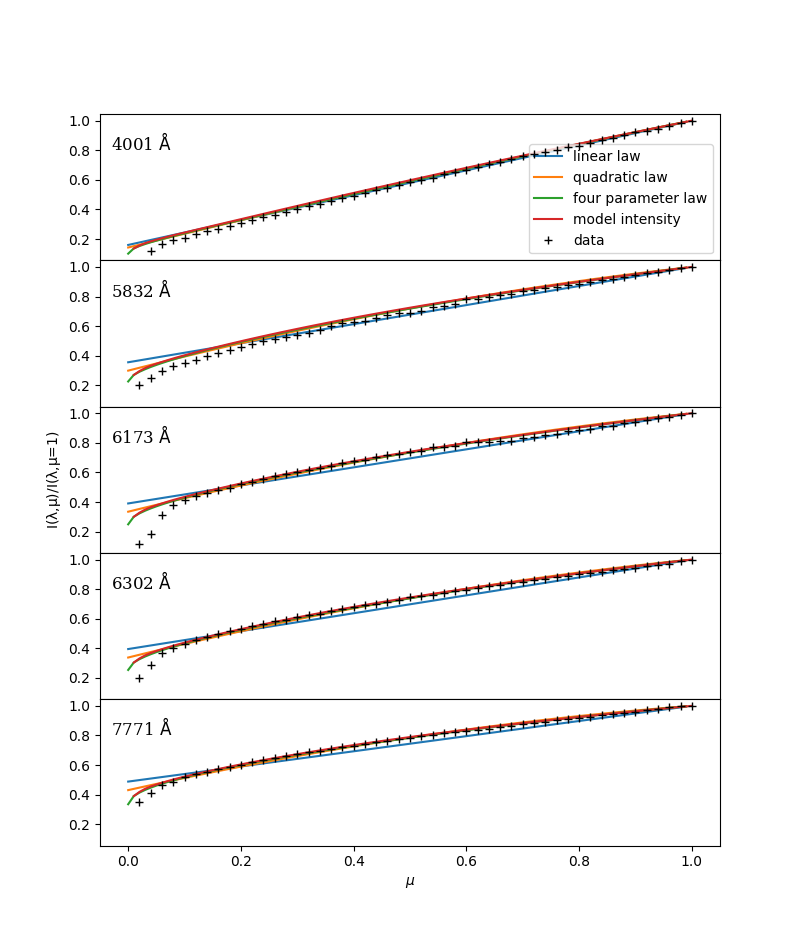}
          \caption{Limb darkening as a function of $\mu$ compared to the intensity variation generated for the Sun from the ATLAS stellar model (red) and the resulting linear (blue), quadratic (orange), and four parameter (green) limb darkening laws. The model and all three laws generally match the solar limb darkening out to $\mu=0.1$ at 4001-6301\,\AA. The linear law is the most different from the data at all wavelengths due to its simplicity.}
          \label{fig:continuumstars}
\end{figure}

Besides the \citet{Neckel94} atlas, there are also several local thermodynamic equilibrium (LTE) codes that model limb darkening curves for a wide range of stars, including the Sun. These limb darkening curves can have a strong impact on stellar and exoplanet characterization \citep[e.g.,][]{Mandel2002,Charbonneau2002,Spake2018}. In order to examine this assumption, we compare our data to three limb darkening approximations obtained by fitting the ATLAS stellar models\footnote{\url{http://kurucz.harvard.edu/grids.html}} using the LDCU python package\footnote{\url{https://github.com/delinea/LDCU}} \citep{Espinoza2015} and assuming an effective temperature of 5778$\pm$58\,K, a log(g) of 4.374$\pm$0.0005, and a turbulent velocity of 1.5$\pm$0.5\,km\,s$^{-1}$ \citep{Gray2008}. The simplest of these approximations is the linear limb darkening law \citep{Milne1921}: 
\begin{ceqn}
\begin{align}
    \frac{I(\lambda,\mu)}{I(\lambda,\mu=1)} = 1 - u(1-\mu),
\end{align}
\end{ceqn}
where $I$ is the intensity as a function of $\lambda$ and $\mu$ and $u$ is the linear limb darkening coefficient. However, linear approximations tend to be too simplistic and cannot reproduce the behavior at the edge of the disk. The four parameter law \citep{Claret2000} is generally considered the most accurate and a good description of the stellar limb darkening and is as follows:

\begin{ceqn}
    \begin{align}
        \frac{I(\lambda,\mu)}{I(\lambda,\mu=1)} = 1 - \sum_{i=1}^4 c_i  (1 - \mu^{i/2}).
    \end{align}
\end{ceqn}

Here $c_i$ represents the limb darkening coefficients. Unfortunately, the quality of the available transit data usually make a four-parameter limb darkening fit degenerate and this likely results in overfitting. For this reason, the squared law is the most widely used \citep{Diaz1992}: 

\begin{ceqn}
    \begin{align}
        \frac{I(\lambda,\mu)}{I(\lambda,\mu=1)} = 1 - c_1(1-\mu) - c_2(1-\mu)^2.
    \end{align}
\end{ceqn}

This law is generally considered capable of reproducing the shape of the CLV without being likely to result in overfitting. 

Figure \ref{fig:continuumstars} shows the trends obtained from the ATLAS model overplotted on the data. It can be seen that while the CLV profiles calculated by the model (shown in red) are accurate within 4\% of the disk-center intensity up to $\mu=0.08$, with the largest discrepancy occurring at 5832Å towards the limb. It can also be seen that, as expected, the four parameter law most closely follows the calculated profile and the data, while the linear fit provides the least accurate representation of the intensity variations. 

Analysis of the coefficients retrieved from the similarly widely used PHOENIX model \citep{Husser2013} shows that the limb darkening is consistent with both the ATLAS model and the data within 4\% of disk-center intensity out to $\mu=0.08$. Overall, it appears that the model parameters obtained from both the ATLAS and PHOENIX codes are in good agreement with solar data at most of the observed continuum wavelengths, although further investigation at longer wavelengths is required.

\begin{figure}
          \centering
          \includegraphics[width=\columnwidth]{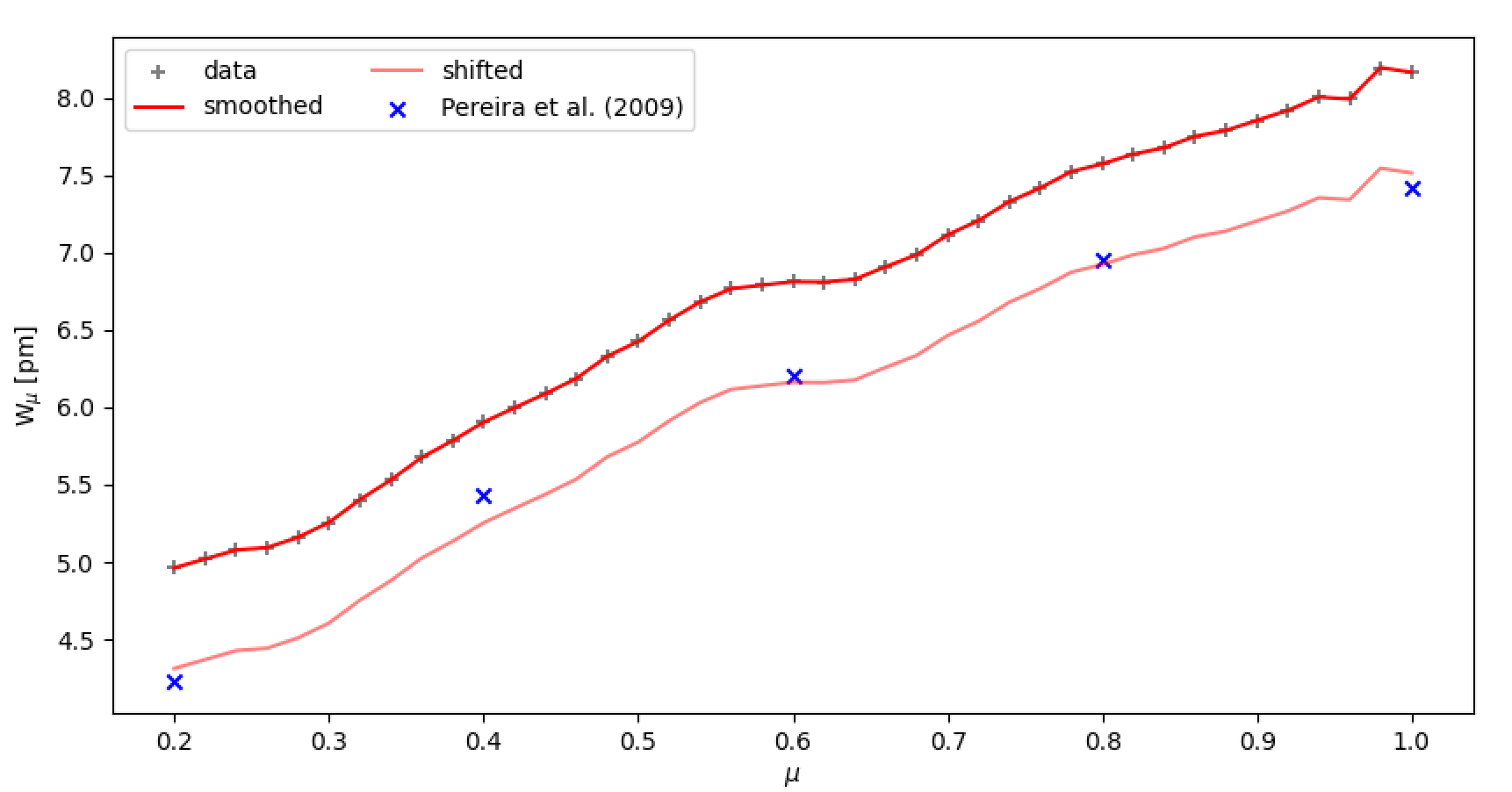}
          \caption{Enlarged version of the first panel of Fig.~\ref{fig:O7773} showing the equivalent width of the O~I~7772~\AA\ line as a function of $\mu$ compared with measured values by \citet{Tiago2009}. We plot the data (gray), the smoothed data (red), the measurements of \citet{Tiago2009} (blue), and our smoothed data shifted down to match the disk center observation of the blue points in light red. The latter was added to ease comparison in regards to the gradient between the two data sets.}
          \label{fig:oxew}
\end{figure}

\subsection{Narrow line parameters}

\begin{figure*}[h!]
          \centering
          \includegraphics[height=0.43\textheight, trim={1.8cm 1cm 2cm 2cm},clip]{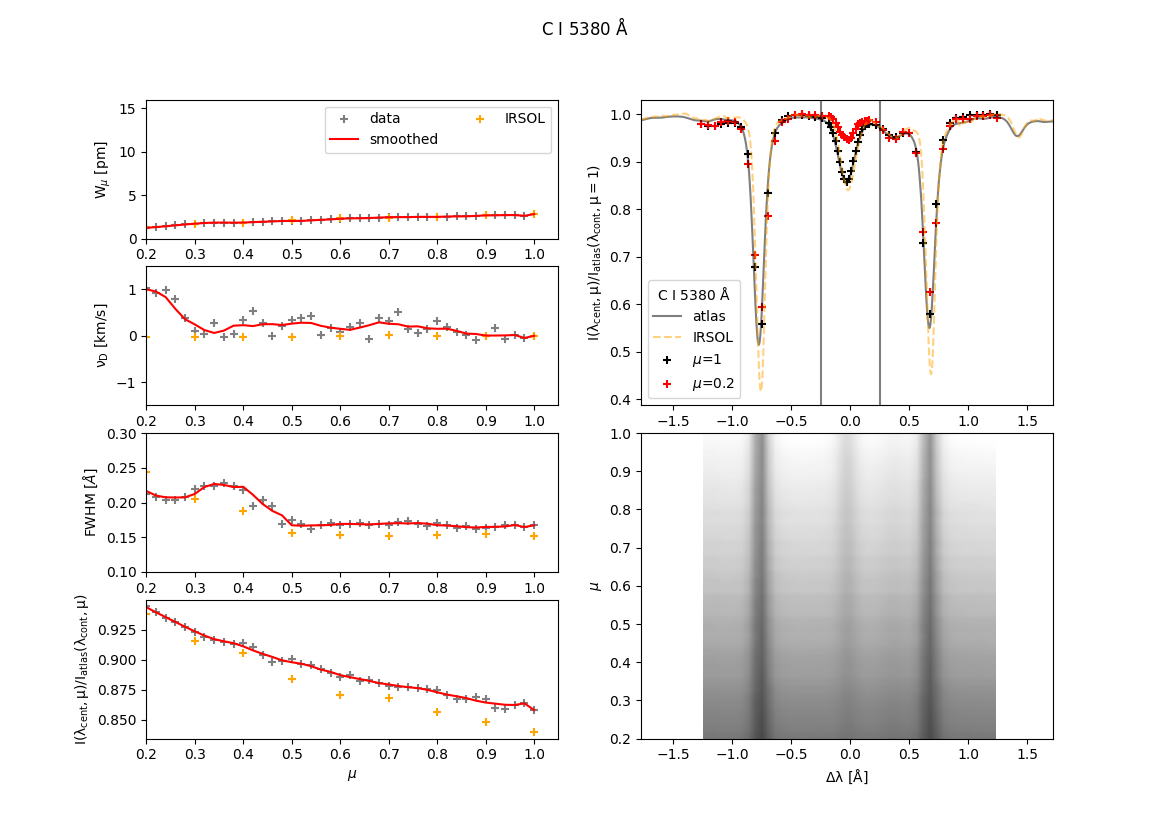}
          \caption{Summary of the CLV data of the C~I~5380~\AA line. \textbf{Left:} Four panels showing the equivalent width, line shift, FWHM, and line depth, respectively. \textbf{Right:} Representative profiles of C~I~5380~\AA. In the upper panel, we plot a convolved profile from the solar atlas as well as our data at $\mu=1$ and $\mu=0.2$. The pair of black lines show the range over which the equivalent width has been calculated. The lower panel shows a 2D representation of all 50 profiles.}
          \label{fig:C5380}

          \centering
          \includegraphics[height=0.43\textheight, trim={2cm 1cm 2cm 2cm},clip]{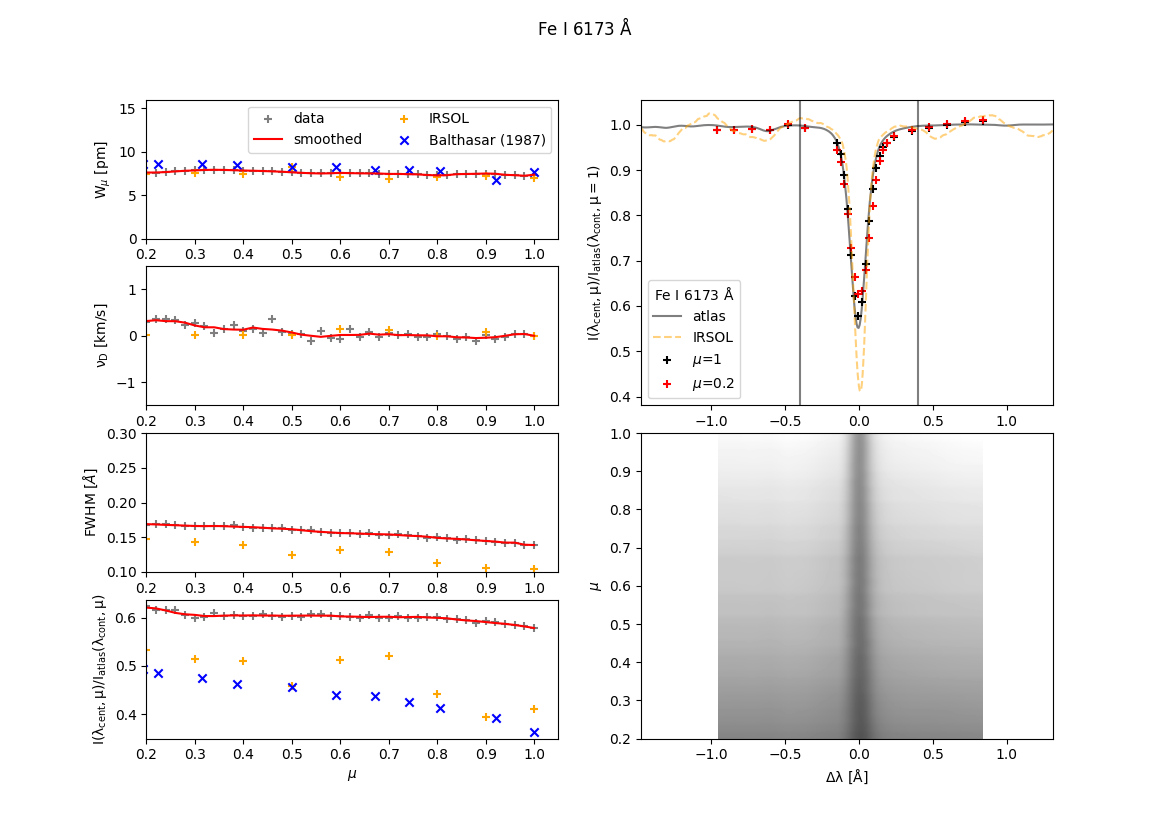}
          \caption{Same as Fig. \ref{fig:C5380}, but for Fe I 6173\, \AA.}
          \label{fig:Fe6173}
\end{figure*}

\begin{figure*}[h!]
          \centering
          \includegraphics[height=0.45\textheight, trim={2cm 1cm 2cm 2cm},clip]{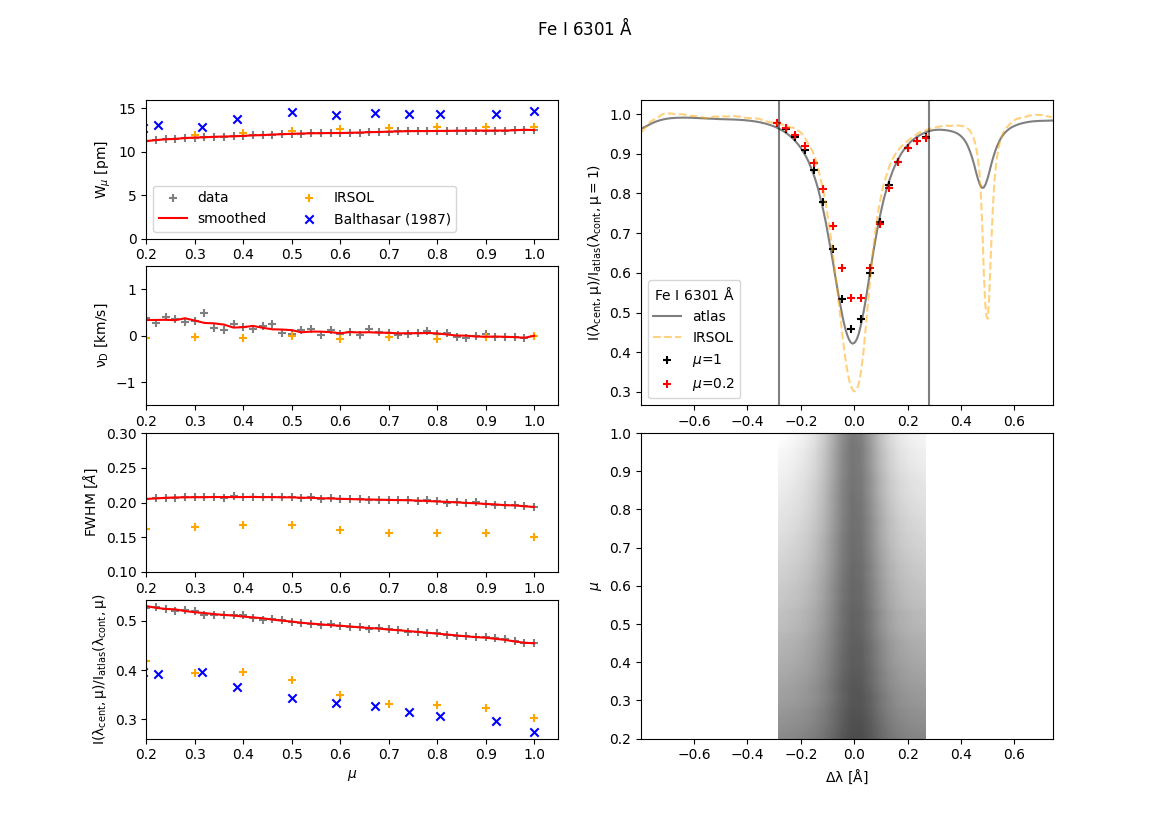}
          \caption{Same as Fig. \ref{fig:C5380}, but for Fe I 6301\, \AA. }
          \label{fig:Fe6302}

          \centering
          \includegraphics[height=0.45\textheight, trim={2cm 1cm 2cm 2cm},clip]{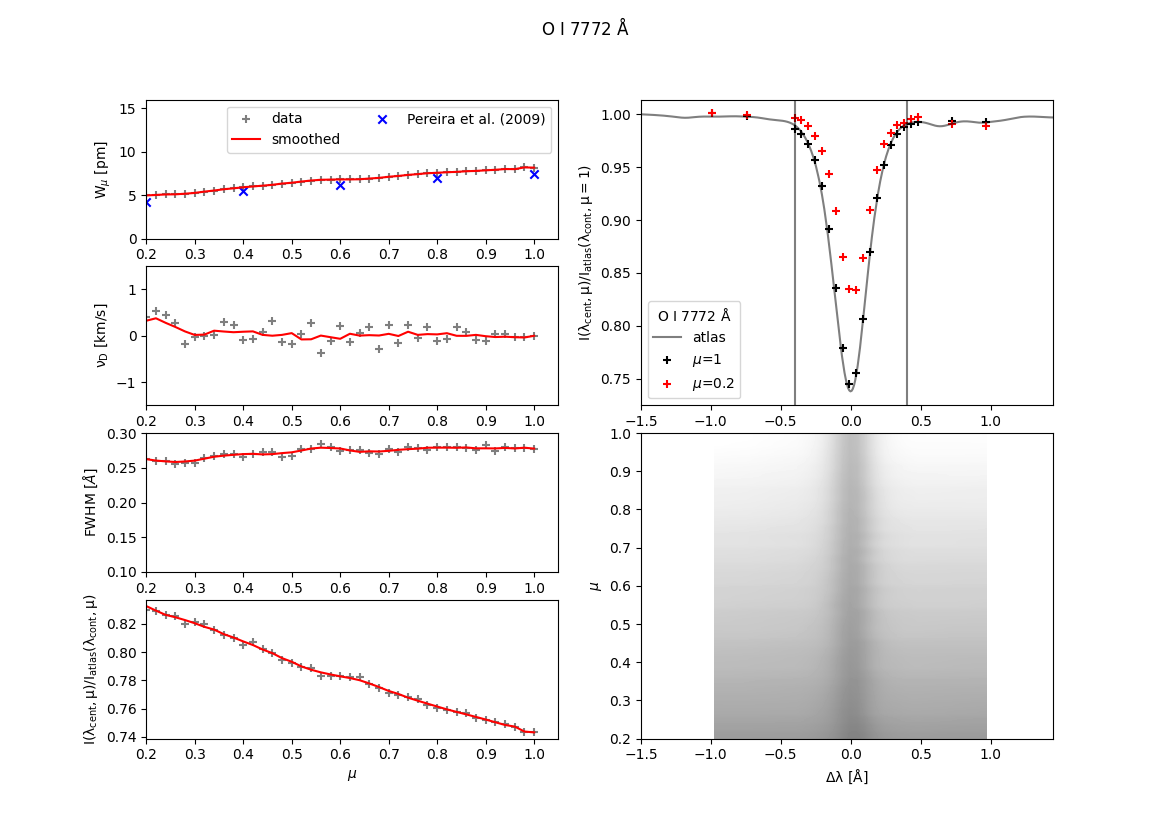}
          \caption{Same as Fig. \ref{fig:C5380}, but for O~I~7772~\AA. Additionally, we include the measured equivalent widths by \citet{Tiago2009} in the upper left panel.}
          \label{fig:O7773}
\end{figure*}

\begin{figure*}[h!]
          \centering
          \includegraphics[height=0.43\textheight, trim={2cm 1cm 2cm 2cm},clip]{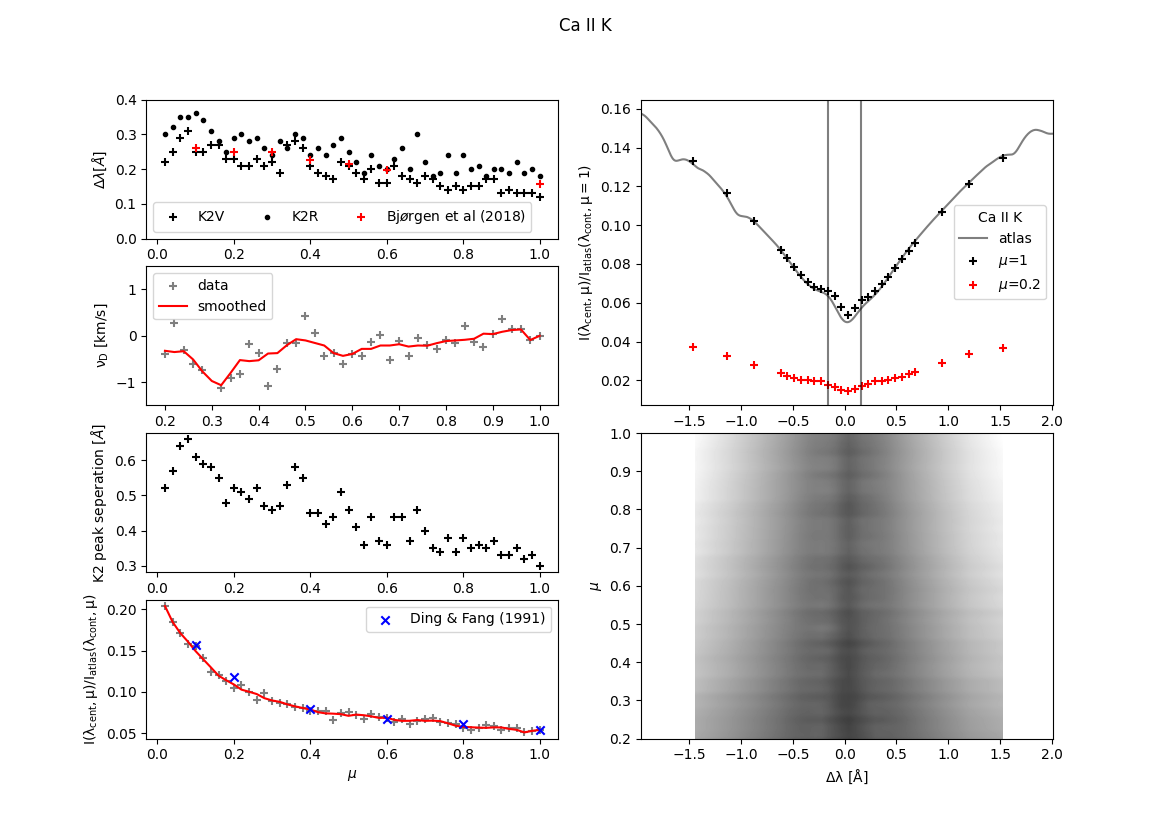}
          \caption{Left: Four panels showing the K2 peak location, line shift, K2 peak separation, and line depth, respectively. Right: Representative profiles of \CaIIK. In the upper panel, we plot a convolved profile from the solar atlas as well as our data at $\mu=1$ and $\mu=0.2$. The lower panel shows a 2D representation of all 50 profiles. }
          \label{fig:Ca3934}

          \centering
          \includegraphics[height=0.43\textheight, trim={2cm 1cm 2cm 2cm},clip]{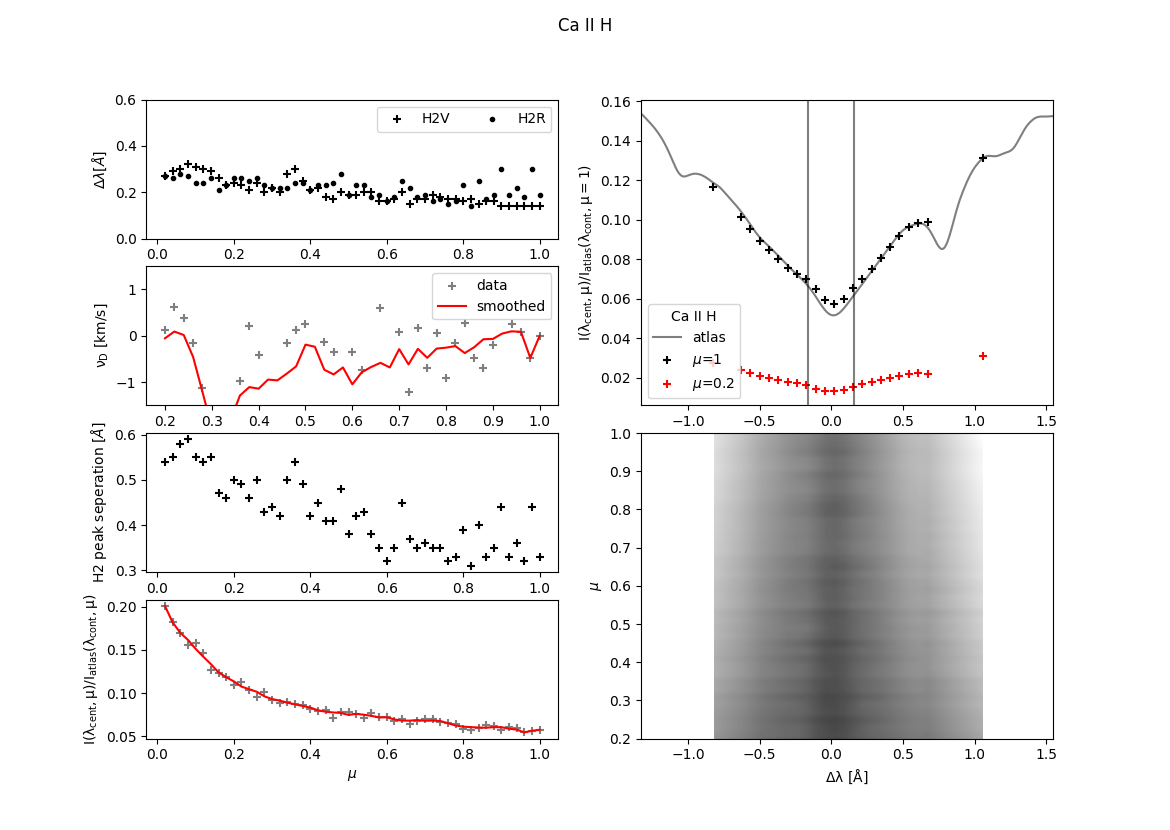}
          \caption{Same as Fig. \ref{fig:Ca3934}, but for \CaIIH. }
          \label{fig:Ca368}
\end{figure*}

\begin{figure*}[h!]
          \centering
          \includegraphics[height=0.43\textheight, trim={2cm 1cm 2cm 2cm},clip]{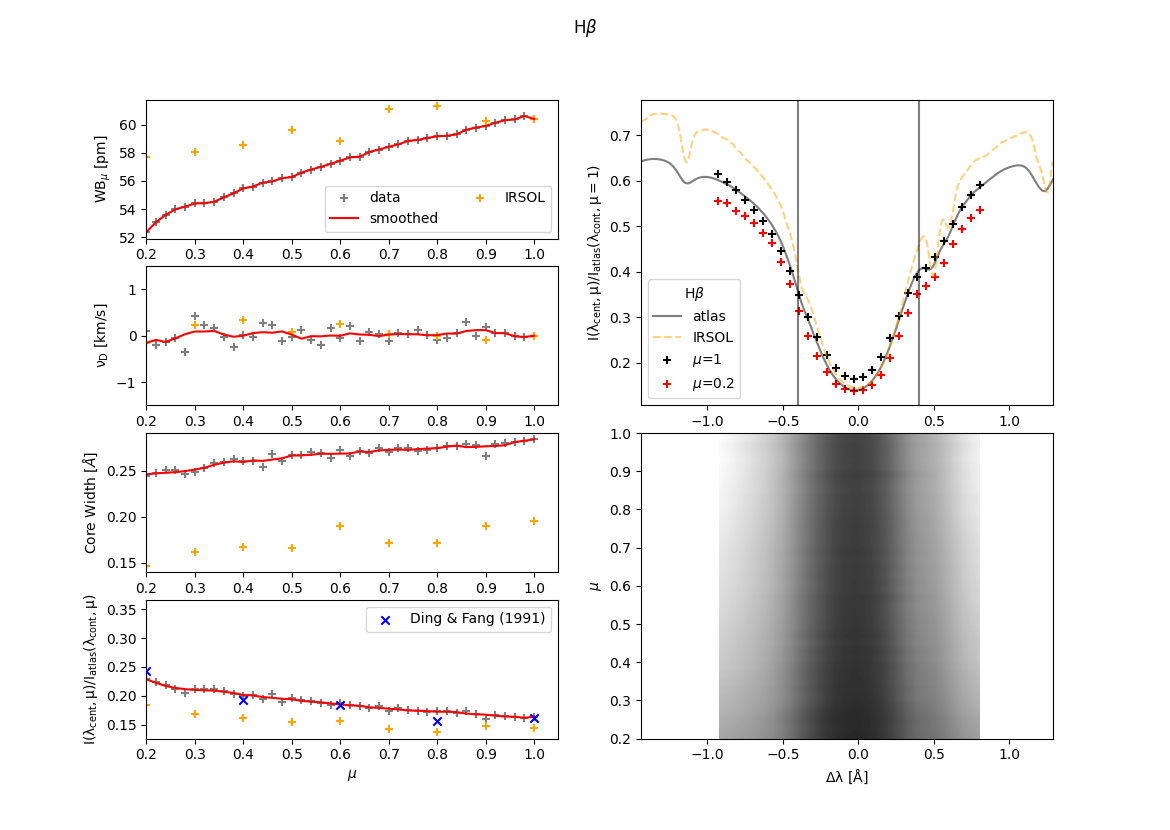}
          \caption{Left: Four panels showing the bound equivalent width, line shift, core width, and line depth, respectively. Right: Representative profiles of H$\beta$. In the upper panel, we plot a convolved profile from the solar atlas as well as our data at $\mu=1$ and $\mu=0.2$. The lower panel shows a 2D representation of all 50 profiles.}
          \label{fig:hbeta}

          \centering
          \includegraphics[height=0.43\textheight, trim={2cm 1cm 2cm 2cm},clip]{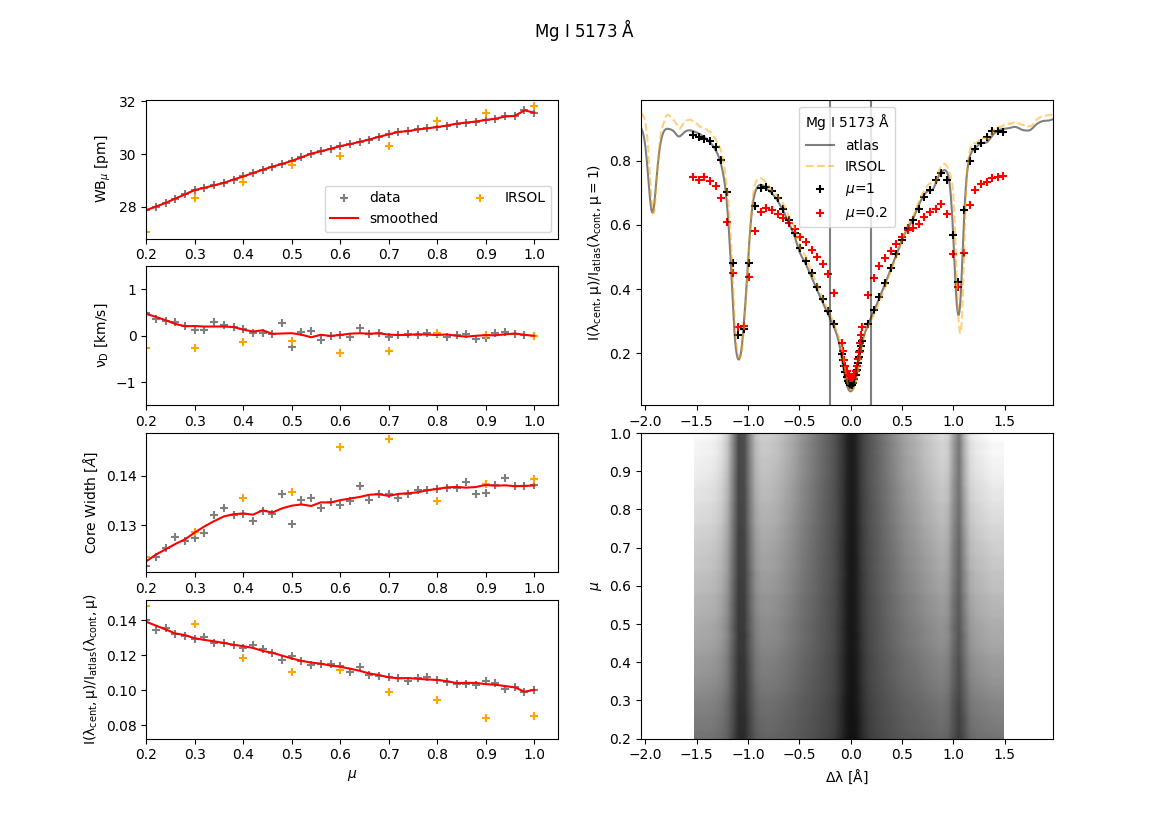}
          \caption{Same as Fig. \ref{fig:hbeta}, but for Mg I 5173 \AA. }
          \label{fig:Mg5173}
\end{figure*}

\begin{figure*}[h!]
          \centering
          \includegraphics[height=0.43\textheight, trim={2cm 1cm 2cm 2cm},clip]{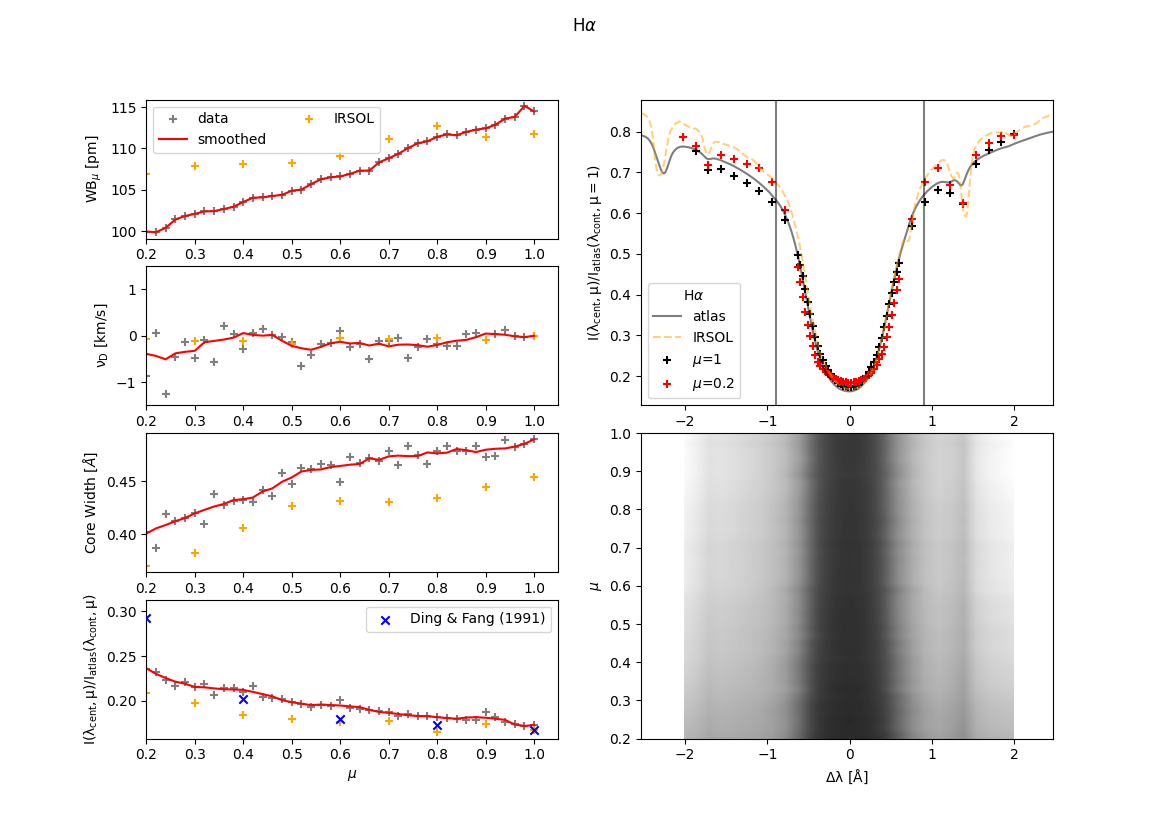}
          \caption{Same as Fig. \ref{fig:hbeta}, but for H$\alpha$. }
          \label{fig:halpha}

          \centering
          \includegraphics[height=0.43\textheight, trim={2cm 1cm 2cm 2cm},clip]{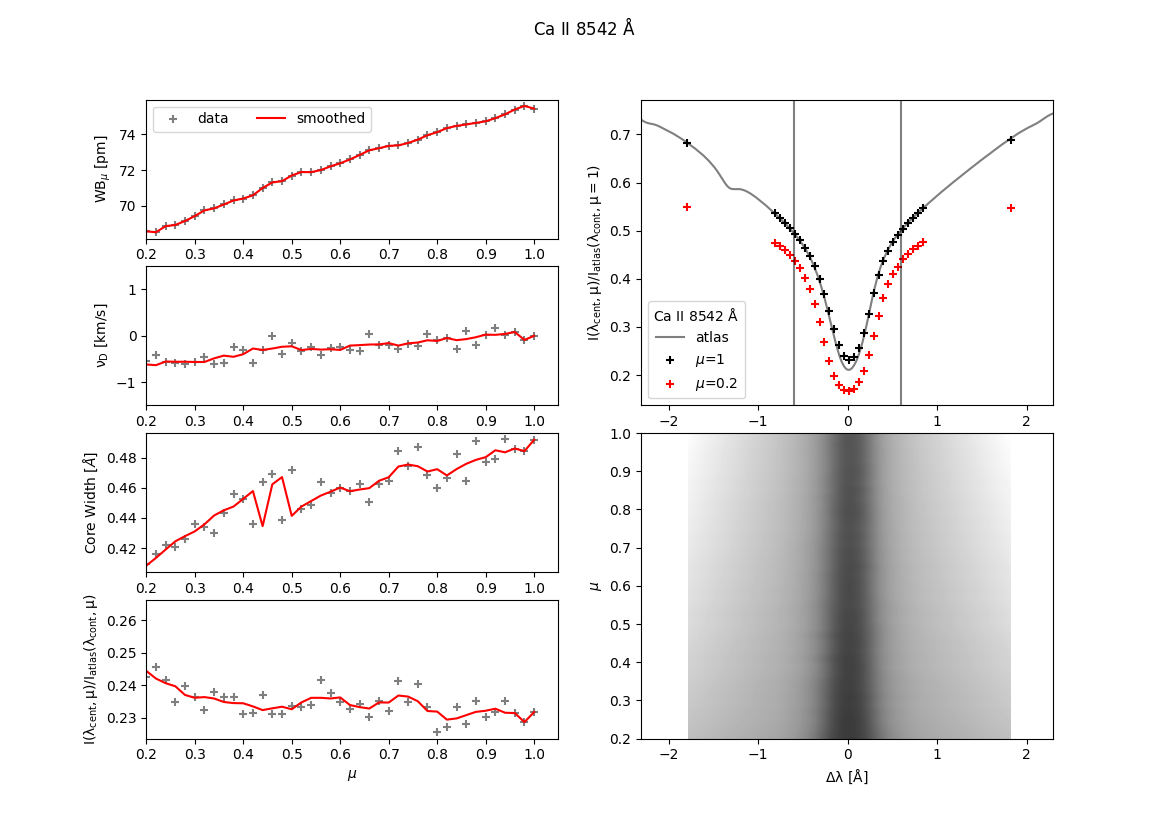}
          \caption{Same as Fig. \ref{fig:hbeta}, but for \cair. }
          \label{fig:CA8542}
\end{figure*}

We present the narrow spectral lines in Figs.~\ref{fig:C5380} to \ref{fig:O7773}. Each figure consists of a set of six panels. The first four on the left side represent the equivalent width, line shift, FWHM, and line depth, respectively. The upper right panel shows a solar atlas profile that is convolved\footnote{The CRISP and CHROMIS instrumental profiles can be calculated with the "spec" module of ISPy \citep{ISPy2021}. However, for convenience, we have added precomputed instrumental profiles to the supplemental material.} with the SST instrumental profile, as well as the average profile at $\mu$=1 and $\mu$=0.2. The last panel shows a 2D representation of the average profiles as a function of $\mu$. All of these lines, except for O~I~7772~\AA\ fall within the spectral range of the IRSOL SS3 spectrum \citep{ramelli17,ramelli19}. In these cases, we have used this data alongside our own for comparison. However, no degradation was applied to the IRSOL SS3 spectra to compare the parameters under a different spectral resolution.

C I 5380 \AA \, ,shown in Fig.~\ref{fig:C5380}, is a relatively weak line that has been shown to virtually disappear at the far limb \citep{Livingston77}. Our observations agree with this statement as we see that the line disappears almost completely around $\mu$=0.1. We find that our equivalent widths perfectly match those of the IRSOL SS3 atlas. However, this is not the case for the line shift. Here we see that, unlike the IRSOL data, our data exhibit a sharp increase around $\mu=0.35$. A similar discrepancy can be seen in the FWHM, where our data exhibit a peak around 0.35\,\AA. After inspecting the average profiles, we find that they take on an asymmetrical shape around $\mu=0.3$, but the reason for this is not clear as the mosaic does not seem to contain more activity in this region. A similar effect is discussed in \citet{Osipov19} where a series of iron lines are observed to have an increasing equivalent width closer to the disk center, but then they start to decrease when very close to the limb. Regardless, additional observations of this $\mu$ range are required. We also see the Fe~I~5379~\AA\ and Ti~II~5381~\AA\ lines in this plot. 

Fe I 6173 \AA\, ,shown in Fig.~\ref{fig:Fe6173}, is a strong iron line, commonly observed by SDO/HMI \citep{SDO2012}. We once again find the equivalent widths of our data and that of the IRSOL SS3 atlas and \citet{horst88} to be in agreement, while the line shift diverges slightly at lower $\mu$. The differences in FWHM and line depth are caused by the different spectral resolutions of the compared data sets.

Fe~I~6301 \AA \,  ,shown in Fig.~\ref{fig:Fe6302}, is a commonly used photospheric line at the SST \citep[e.g.,][]{2021A&A...647A.188D} and Hinode/SP \citep{Kosugi2007}. Here we find that our observations match those of the IRSOL SS3 atlas for the equivalent width and line shift, but both data sets return equivalent width values that are lower than those reported in \citet{horst88}. This is likely due to the fact that our data do not include enough of a continuum. We again note the same divergence for the line shift as was seen in the previous two lines. 

O\,I\,\,7772 \AA\, ,shown in Fig.~\ref{fig:O7773},  is a strong oxygen line often used for oxygen abundance measurements \citep{2021MNRAS.508.2236B}. This line falls outside of the wavelength range of the IRSOL SS3 atlas. We compared our values to the equivalent width measurements presented in \citet{Tiago2009}. We find our $W_\mu$ values to be systematically above those of \citet{Tiago2009}, but we find a similar gradient in the line (see Fig.~\ref{fig:oxew}). The discrepancy in values is likely once again caused by a difference in the definition of the continuum and the bounds between which the equivalent width is calculated.

\begin{table}[]
\centering
\caption{Separation parameters for broad lines.}
\begin{tabular}{ll}
\hline
\hline
Line & $\Delta\lambda_l$ {[}m\AA{]} \\ \hline
H$\beta$ 4861 \AA     & 400                              \\ 
Mg I 5173 \AA & 200\\
\halpha 6563 \AA     & 900                              \\ 
\cair &                    600               \\ \hline
\end{tabular}\label{tab:width}

\end{table}

\subsection{Broad line parameters}\label{broadsec}

The \CaIIK line, shown in Fig.~\ref{fig:Ca3934}, has a more complicated shape compared with most of the other lines that we study. This is because the line source function decouples from temperature just after the temperature minimum, which results in two peaks on either side of the line core. The blue peak (K2V) is sensitive to a larger height range than the red peak (K2R) \citep[e.g.,][]{johan18}. In the quiet Sun, we find that this often leads to the K2R peak disappearing, with about 25\% of quiet-Sun \CaIIK profiles being reversal-free \citep{Rezaei18}. This effect is strengthened when averaging, as the already low intensity of the K2R peak gets smeared out. We can see this in the solar atlas (upper right panel in Fig.~\ref{fig:Ca3934}), where there is no clear K2R peak, while a K2V peak is visible. The same is true for our data at $\mu=1$. However, the location of both peaks can still be inferred from the wavelength-dependent standard deviation of each averaged profile\footnote{The standard deviation of the intensity of the individual pixels inside each $\mu$ bin for each wavelength point in the spectral line.}, which is largest at the location of the peaks. This location is plotted in the top left panel of the same figure. Additionally, we have overplotted observational measurements of the K2R peak location from \citet{johan18} in the same panel for comparison. While our data are noisier than the literature values, they match the trend for almost the entire range apart from the last few points at the extreme limb, where the peaks seem to be moving closer together again. For the line depth, we have overplotted values from \cite{Ding91}, which match very well with our data. In Fig.~\ref{fig:chrom} we can see that apart from the noise, the entire profile has a similar behavior up until $\mu = 0.2$, at which point the line wings dip down stronger. This is in line with the derived parameters discussed above, where most of the change in the profile seems to come from shifts in the K2 peaks.

\CaIIH, shown in Fig.~\ref{fig:Ca368}, behaves very similarly to \CaIIK. We find a lower spread in the R and V peak shifts than we did for \CaIIK, but a similar behavior with regard to both peaks moving apart with lower $\mu$ until the last few measurements. We find that the line core behaves in a similar way to \CaIIK, and that the velocities seem to be slewing toward the blue. In Fig. \ref{fig:chrom} we also see a similar behavior to that of \CaIIK, where most of the profile seems to undergo the same limb darkening effects until roughly $\mu$=0.2

H$\beta$, shown in Fig.~\ref{fig:hbeta},  differs strongly from the IRSOL SS3 atlas values for the bound equivalent width. This is likely due to the chromium blend that can be found in the blue wing of the line, which has a differing depth due to the mismatch in spectral resolution between the two data sets. Nevertheless, our data once more are in strong agreement with the values provided by \citet{Ding91}. In Fig.~\ref{fig:chrom} we see a slightly larger spread in the limb darkening curves, but with largely the same shape. The line core seems to be dipping slightly faster than the wings.  

Mg I 5173 \AA \, shown in Fig.~\ref{fig:Mg5173}, is a line that forms around the temperature minimum. When comparing our data to the IRSOL SS3 atlas, we find that the bound equivalent width matches the IRSOL values closely. This is not the case for the line shift, where we see a curve toward the red at low $\mu$ values similar to the behavior of the photospheric lines. In Fig.~\ref{fig:chrom} we see a much bigger spread in the limb darkening curves, with the far wings following the behavior of the continuum, and the core intensities being less affected. This data set also contains the Fe~I~5172~\AA\ and the Ti~I~5174~\AA\ lines.   

H$\alpha$, shown in Fig.~\ref{fig:halpha}, has similar behavior to what we saw in the H$\beta$ line, where our data strongly differ from the IRSOL data. The difference in trend between our data and that of IRSOL is likely due to the telluric water line that is visible in the red wing of the IRSOL data, but it is completely washed out in ours. The line depth has a different gradient to that of the IRSOL data, but it matches the data from \citet{Ding91} reasonably well, except for the last point. In Fig.~\ref{fig:chrom} we see that the limb darkening behavior of this line is quite different from H$\beta$. First, there are the wing points that dip strongly, but these can be explained by the fact that H$\alpha$ samples further into the far wings. Second, the line core is more affected by the limb darkening than the near wings and thus acting oppositely to H$\beta$. Finally, there is a larger spread in the profiles' limb darkening curves.

Ca~II~8542~\AA \, shown in Fig~\ref{fig:CA8542},  behaves similarly to the other chromospheric lines when it comes to the studied parameters. In Fig. \ref{fig:chrom} we see \cair behaves in a similar manner to H$\beta$, with the line core being more affected by the limb darkening than the continuum. However, besides that, we see that the normalized line core closely follows the continuum curve until roughly $\mu$=0.3.

Finally, we would like to point to the fact that all of our narrow photospheric lines as well as Mg~I~5173~\AA\ show a positive (red) shift of their core with decreasing $\mu$. This can be explained by the fact that the disk center has a blue-shifted velocity on average because of granular motions. This average blue shift decreases with $\mu$ as we start to look at the granules from the side. However, we find that all of the remaining broad chromospheric lines tend to shift toward the blue with decreasing $\mu$. It is possible that this is because the chromospheric canopy tends to be horizontal and that these velocities enter our line of sight at higher $\mu$ angles. The implied preferential direction of the canopy could be explained by the projection effects of a radial structure. At lower $\mu$ the part pointing toward the disk center looks bigger than the part pointing away from it, which would result in a net blue shift in the line of sight. The fact that Mg I 5173 \AA\ behaves more like a photospheric line is in line with this given that we do not see horizontal canopy structures in the line core. However, it is also possible that this is caused by an unforeseen bias in the averaging process, and further investigation into individual profile statistics is required.

\section{Conclusions}\label{conclusions}

In this work, we present the results derived from a series of high-resolution mosaics of the broad \CaIIHK Mg~I~5173~\AA, \Hbeta, \halpha, and \cair lines, as well as of the four narrow lines C~I~5380~\AA, Fe~I~6173~\AA, \FeI, and O~I~7772~\AA\. These mosaics, composed of 25 pointings, each spanning roughly 60\,$\times$\,60 arcseconds, allowed us to create a series of 50 high-resolution reference profiles, spaced by 0.02 in the $\mu$ scale for each of the observed lines. Additionally, we created a second set of reference profiles that were smoothed using a box smoothing algorithm in order to remove imprints caused by the 5-minute oscillations.

We tested the accuracy of the limb darkening curves obtained from these reference profiles against CLVs of atlas continuum intensities and synthetic continuum intensities, and found it to match in all cases (see Figs.~\ref{fig:continuum} and \ref{fig:continuumstars}). In fact, we were able to extend the continuum limb darkening curves of selected wavelengths past the $\mu = 0.12$ cutoff that was imposed by \citet{Neckel94}. We compared our continuum curves to stellar limb darkening models and found a generally good agreement above $\mu$=0.08, where our data become less reliable. Of the most commonly used limb darkening laws, the quadratic and four parameter laws showed the best match, while the linear law deviated the most due to its simplicity. These curves are available in the supplemental material.

We then investigated the narrow lines by measuring the equivalent width, line shift, FWHM, and line depth of each of these lines, as well as by comparing these parameters to existing literature where this was available. In general, we find that our data match existing literature values, and that discrepancies can be explained by differences in the continuum definition and integration range for the equivalent width.

The same was done for the regularly shaped broad lines, with a slight modification to the definition of the equivalent width and FWHM, following conventions started by \citet{2009cauzzi}. This allowed us to study the chromospheric behavior of these lines without it being drowned out by their large photospheric wings, but such a method is very sensitive to the spectral resolution of the studied line. For this reason we suggest the future comparisons using this parameter be done on lines with the same spectral resolution. We find that these chromospheric lines exhibit a small shift that tends toward the blue, a phenomenon that could potentially be caused by projection effects of the chromospheric canopy which tends to flow horizontally. We note that Mg~I~5173~\AA\ tends to behave much more like a photospheric line with a line shift toward the red. This could be due to the low formation height of the line, where it is still influenced by the granular velocities.  

For the \CaIIHK lines, we investigated the separation between the reversal peaks on both ends of the line core. These peaks are not always visible in our spectra but they were found by looking at the wavelength-dependent standard deviation. In both cases, we find that the peaks move away from one another with decreasing $\mu$, and that our values match the observational values presented in \citet{johan18}. As with the other broad lines, we find the \CaIIHK lines tend to shift toward the blue for decreasing $\mu$.


For future work of this kind, we would advise taking longer time series at each mosaic position or taking multiple time-independent mosaics of each line in order to average out the p modes and other local effects, since this is the main source of uncertainty in our data. This would come at the cost of temporal coherence between overlapping mosaic parts, which illustrates the trade-offs that one must consider while making these kinds of observations. A potential way around this would be to apply an integral field spectrometer that can simultaneously capture spatial and spectral information and to drift scan this instrument across the Sun, as was done with single wavelengths. However, the FOV of these kinds of instruments tends to be rather small, which will provide us with fewer pixels to average over. 

We hope that these average profiles can be of use to the community for abundance studies and as an incident boundary condition for radiative transfer modelers. We also aim to apply these profiles to improve the intensity calibrations of the SSTRED pipeline.

\begin{acknowledgements}
We are grateful to Sepideh Kianfar and Carolina Robustini for participating in the SST observations and to Mats Löfdahl for the stimulating discussions on this topic. We also thank Horst Balthasar and the anonymous referee for their valuable suggestions during the peer-review process.
This work was supported through the CHROMATIC project (2016.0019) of the Knut and Alice Wallenberg foundation and by the European Commission’s Horizon 2020 Program under grant agreements 824064 (ESCAPE -- European Science Cluster of Astronomy \& Particle Physics ESFRI Research Infrastructures) and 824135 (SOLARNET -- Integrating High Resolution Solar Physics).
This project has received funding from the European Research Council (ERC) under the European Union's Horizon 2020 research and innovation program (SUNMAG, grant agreement 759548). This research is supported by the Research Council of Norway, project number 325491 and through its Centers of Excellence scheme, project number 262622.
Contributions by D.P have been carried out within the framework of the NCCR PlanetS supported by the Swiss National Science Foundation under grants 51NF40\_182901 and 51NF40\_205606. D.P also acknowledges support of the Swiss National Science Foundation under grant number PCEFP2\_194576.
The Swedish 1-m Solar Telescope is operated on the island of La Palma by the Institute for Solar Physics of Stockholm University in the Spanish Observatorio del Roque de los Muchachos of the Instituto de Astrof\'isica de Canarias. The Institute for Solar Physics was supported by a grant for research infrastructures of national importance from the Swedish Research Council (registration number 2017-00625).
This research has made use of NASA's Astrophysics Data System Bibliographic Services. 
We acknowledge the community effort devoted to the development of the following open-source packages that were used in this work: numpy (numpy.org), matplotlib (matplotlib.org), astropy (astropy.org).

We extensively used the CRISPEX analysis tool \citep{Gregal12}, the ISPy library \citep{ISPy2021}, Specutils \citep{Specutils22}, sTools \citep{stools}, and SOAImage DS9 \citep{2003DS9} for data reduction and visualization during the preparation of this article, as well as Google Translate and DeepL.com for translating non-English articles.

\end{acknowledgements}

\bibliographystyle{aa}
\bibliography{ref}

\end{document}